\documentclass[acmsmall,nonacm,screen]{acmart}

\AtBeginDocument{%
  }

\setcopyright{acmlicensed}
\copyrightyear{2018}
\acmYear{2018}
\acmDOI{XXXXXXX.XXXXXXX}
\acmConference[Conference acronym 'XX]{Make sure to enter the correct
  conference title from your rights confirmation email}{June 03--05,
  2018}{Woodstock, NY}
\acmISBN{978-1-4503-XXXX-X/2018/06}

\usepackage{algorithmic}
\usepackage[ruled,linesnumbered,vlined]{algorithm2e}
\usepackage{multicol}
\usepackage{graphicx}
\usepackage{textcomp}
\usepackage{color, colortbl}
\usepackage{xcolor}
\PassOptionsToPackage{prologue, table}{xcolor}
\usepackage[nounderscore]{syntax}
\usepackage{microtype}
\usepackage{url}
\usepackage{xspace}
\usepackage{booktabs, multirow} 
\usepackage{listings}
\usepackage{tikz}
\usepackage{mdframed} 
\usepackage{caption}
\usepackage{subcaption}
\usepackage[normalem]{ulem}
\usepackage{diagbox}
\usepackage{cleveref}
\usepackage{subcaption}
\usepackage{soul}
\usepackage[normalem]{ulem}

\usepackage{savesym}
\savesymbol{colonapprox}
\savesymbol{colonsim}
\savesymbol{Bbbk}
\usepackage{amssymb,amsthm,amsmath,stmaryrd}
\usepackage{mathtools}
\usepackage{mathpartir}
\usepackage{colonequals}
\usepackage{cleveref}
\usepackage{float}
\usepackage{array} 

\newtheorem*{theorem*}{Theorem}

\newcommand{\ignore}[1]{}

\begin{document}


\newcommand{\name}{Mapple\xspace}

\newcommand{\anjiang}[1]{\textcolor{blue}{#1}}

\newcommand{\ke}[1]{\textcolor{red}{#1}}

\newcommand{\hang}[1]{\textcolor{blue}{#1}}

\newcommand{\locreduction}{14$\times$\xspace}
\newcommand{\mapperspeedup}{1.34$\times$\xspace}
\newcommand{\primitivespeedup}{1.83$\times$\xspace}
\newcommand{\slowdownmax}{3.5$\times$\xspace}

\newcommand{\improveavg}{16\%\xspace}
\newcommand{\improvemax}{83\%\xspace}

\newcommand{\CodeIn}[1]{\texttt{#1}}

\newcommand{\ccc}[1]{\hfill \begin{small}\# #1\end{small}}
\newcommand{\parabf}[1]{\noindent\textbf{#1}}

\newcommand{\legion}{Legion\xspace}
\newcommand{\regent}{Regent\xspace}
\newcommand{\charmpp}{Charm++\xspace}
\newcommand{\starpu}{StarPU\xspace}
\newcommand{\parsec}{PaRSEC\xspace}
\newcommand{\chapel}{Chapel\xspace}
\newcommand{\hpx}{HPX\xspace}
\newcommand{\pathways}{Pathways\xspace}
\newcommand{\sequoia}{Sequoia\xspace}
\newcommand{\openmp}{OpenMP\xspace}
\newcommand{\ompss}{OmpSs\xspace}
\newcommand{\ray}{Ray\xspace}
\newcommand{\dask}{Dask\xspace}
\newcommand{\spark}{Spark\xspace}
\newcommand{\mapreduce}{MapReduce\xspace}
\newcommand{\xten}{X10\xspace}
\newcommand{\hypertool}{Hypertool\xspace}
\newcommand{\dague}{DAGuE\xspace}
\newcommand{\distal}{DISTAL\xspace}

\newcommand{\forder}{Fortran-order\xspace}
\newcommand{\corder}{C-order\xspace}
\newcommand{\soa}{struct-of-arrays\xspace}
\newcommand{\aos}{array-of-structs\xspace}

\newcommand{\mergecode}{\CodeIn{merge}\xspace}
\newcommand{\splitcode}{\CodeIn{split}\xspace}
\newcommand{\swapcode}{\CodeIn{swap}\xspace}
\newcommand{\balancesplitcode}{\CodeIn{balance\_split}\xspace}
\newcommand{\autosplitcode}{\CodeIn{decompose}\xspace}

\newcommand{\decompose}{\texttt{decompose}\xspace}

\newcommand{\mergenocode}{merge\xspace}
\newcommand{\splitnocode}{split\xspace}
\newcommand{\swapnocode}{swap\xspace}
\newcommand{\balancesplitnocode}{balance\_split\xspace}
\newcommand{\autosplitnocode}{decompose\xspace}

\newcommand{\circuit}{Circuit\xspace}
\newcommand{\stencil}{Stencil\xspace}
\newcommand{\pennant}{PENNANT\xspace}
\newcommand{\cannon}{Cannon’s\xspace}
\newcommand{\pumma}{PUMMA\xspace}
\newcommand{\summa}{SUMMA\xspace}
\newcommand{\johnson}{Johnson's\xspace}
\newcommand{\solomonik}{Solomonik's\xspace}
\newcommand{\cosma}{COSMA\xspace}

\newcommand{\kw}[1]{\mathrm{#1}} 

\newcommand{\smallspace}{\hspace{0.5em}}
\newcommand{\precaptionspace}{\vspace{-1.0em}}
\newcommand{\postcaptionspace}{\vspace{-0.6em}}
\newcommand{\pretabcaptionspace}{\vspace*{-1.5ex}}
\newcommand{\posttabcaptionspace}{\vspace*{-1.5ex}}

\lstdefinestyle{taskcode}{
  language=C, 
  basicstyle=\ttfamily\small,
  columns=fullflexible,
  frame=single,
  breaklines=true,
  showstringspaces=false,
  commentstyle=\color{gray}\itshape
}

\definecolor{commentgreen}{RGB}{63,127,95}
\definecolor{dslpurple}{rgb}{0.4, 0.2, 0.6}
\definecolor{dslkeyword}{RGB}{0, 0, 180}

\lstdefinestyle{dslstyle}{
  language=Python,
  basicstyle=\ttfamily\tiny,
  numbers=none,
  numbersep=4pt,
  escapeinside={(*@}{@*)},
  keywordstyle=\color{dslkeyword}\bfseries,
  identifierstyle=\color{black},
  morekeywords={Task,Machine,decompose,IPoint,ISpace,MSpace,Tuple,Region, Layout, GarbageCollect, Backpressure, IndexTaskMap},
  commentstyle=\color{gray},
  showstringspaces=false
}

\lstdefinestyle{cppstyle}{
  language=C++,
  basicstyle=\ttfamily\tiny,
  numbers=none,
  numbersep=4pt,
  escapeinside={(*@}{@*)},
  commentstyle=\color{commentgreen}\itshape,
  keywordstyle=\color{blue}\bfseries,
  stringstyle=\color{orange},
  showstringspaces=false
}

\lstdefinelanguage{Regent}
{morekeywords=[1]{task, reads, while, true, writes, for, do, in, if, else, then, elseif, import, require, regentlib.start, end, var, __demand, __cuda, __inline, __index_launch, __vectorize, fill, struct},
morekeywords=[2]{return, region, fspace, ispace, partition, image, preimage, equal, disjoint, aliased, int, double, int1d, int2d, int3d, break, and},
sensitive=true,
morecomment=[l]{--},
morecomment=[l]{\#},
morecomment=[s]{--[[}{--]]},
morestring=[b]",
}

\definecolor{keywordcolor}{rgb}{0.5,0,0.5}
\definecolor{commentcolor}{rgb}{0.5,0.5,0.5}

\title{Mapple: A Domain-Specific Language for Mapping Distributed Programs}

\author{Anjiang Wei}
\affiliation{%
  \institution{Stanford University}
  \country{USA}
}
\email{anjiang@cs.stanford.edu}

\author{Rohan Yadav}
\affiliation{%
  \institution{Stanford University}
  \country{USA}
}
\email{rohany@cs.stanford.edu}

\author{Hang Song}
\affiliation{%
  \institution{Stanford University}
  \country{USA}
}
\email{songhang@stanford.edu}

\author{Wonchan Lee}
\affiliation{%
  \institution{NVIDIA}
  \country{USA}
}
\email{wonchanl@nvidia.com}

\author{Ke Wang}
\affiliation{%
  \institution{Nanjing University}
  \country{China}
}
\email{kwg@nju.edu.cn}

\author{Alex Aiken}
\affiliation{%
  \institution{Stanford University}
  \country{USA}
}
\email{aiken@cs.stanford.edu}

\begin{abstract}
Optimizing parallel programs for distributed systems is a complex task, often requiring significant code modifications. Task-based programming systems improve modularity by separating performance decisions from application logic, but their mapping interfaces are low-level. We introduce \name{}, a high-level, declarative programming interface for mapping distributed applications. \name{} provides transformation primitives to resolve dimensionality mismatches between task and processor spaces, including a key primitive, \emph{decompose}, that helps minimize communication volume. We implement \name{} on top of the Legion runtime by translating \name{} mappers into its low-level C++ interface. Across nine applications, including six matrix multiplication algorithms and three scientific computing workloads, \name{} reduces mapper code size by \locreduction{} and enables performance improvements of up to \mapperspeedup{} over expert-written C++ mappers. In addition, the \emph{decompose} primitive achieves up to \primitivespeedup improvement over existing dimensionality-resolution heuristics.
\end{abstract}

\maketitle

\section{Introduction}
\label{sec:intro}

Optimizing parallel programs on modern distributed systems remains complex and labor-intensive despite decades of research. Programmers must navigate intricate low-level programming interfaces and often make substantial code modifications to achieve high performance, making tuning challenging even for experts.

Two dominant programming paradigms exist today. The first is the Message Passing Interface (MPI)+X model~\cite{gropp1999using,sanders2010cuda,chandra2001parallel,dagum1998openmp,yang2011hybrid} where MPI is used for distributed-memory parallelism (across computers), and X is typically a shared-memory programming model for intra-node parallelism (within one computer), such as OpenMP or CUDA. This model gives programmers fine-grained control over performance-critical aspects but significantly compromises productivity. Achieving performance gains often requires substantial changes throughout the application code, making it difficult to optimize without inadvertently affecting application logic. The second paradigm is task-based programming systems~\cite{chamberlain2007parallelchapel,augonnet2009starpu,fatahalian2006sequoia,kale1993charm++,duran2011ompss,kaiser2014hpx,slaughter2015regent}, which separate performance decisions from core application logic. Tasks are functions whose dependencies are analyzed by the runtime and scheduled across processors. These systems offer interfaces for specifying performance policies such as task and data placement, allowing developers to tune performance independently of application logic. This separation improves modularity and makes tuning more systematic and less error-prone.

In task-based systems, the programming interface responsible for performance-critical decisions is known as the \emph{mapping} or \emph{scheduling} interface; we will refer to the concrete implementation of these decisions as a \emph{mapper}. A central aspect of mapping is to partition tasks and data and assign them to processors in a distributed system. Distributed algorithms, such as those for matrix multiplication~\cite{cannon1969cellular,van1997summa,choi1994pumma,solomonik2011communication,johnsonagarwal1995three,cosmakwasniewski2019red}, rely on carefully crafted tensor partitioning and processor mappings. For instance, \cannon algorithm~\cite{cannon1969cellular} can experience up to a \slowdownmax slowdown when standard runtime heuristics are used instead of the intended mapping strategy.

A major limitation of existing mapping interface designs is their low-level nature and lack of abstraction, which exposes users to the complexity and intricacies of the underlying runtime system. These interfaces are often tightly coupled with the runtime, making them difficult to use, especially for application developers unfamiliar with runtime internals. For instance, defining a new mapper in the Chapel framework~\cite{chamberlain2007parallelchapel} requires implementing 19 functions according to 33 pages of documentation~\cite{chamberlain2010user,chamberlain2011authoring}; the Legion mapping interface has more user-implementable callbacks \cite{bauer2012legion}.

In this work, we introduce \name{}, a domain-specific language for mapping distributed applications. \name{} provides a high-level abstraction that hides low-level runtime details while still exposing performance-critical mapping decisions. Compared to the original interface, \name{} reduces code size by \locreduction{} without sacrificing performance (see~\Cref{sec:result:locreduction}).

A core challenge in any mapping interface targeting distributed applications is to determine how a multi-dimensional \emph{task space}, defined by the algorithm's set of tasks to execute, maps onto a multi-dimensional \emph{processor space}, defined by the machine's hierarchical architecture (e.g., a cluster of \( n \) nodes with \( m \) GPUs per node). This mapping process is directly responsible for the amount of data movement across processors, which is a key factor in achieving high performance on distributed systems. This challenge is compounded by the fact that these two spaces often differ in dimensionality, making writing an effective mapping inherently complex.

To address potential mismatches in dimensionality between the two spaces, \name{} introduces a set of transformation primitives that operate on the processor space, which raises another question: Can we automatically determine the optimal processor space transformation, and what principles guide the resolution of dimensionality mismatches? To answer these questions, we introduce a new primitive, \emph{decompose}, which provides a unified solution for common cases where communication has locality. Through theoretical analysis, we show that \emph{decompose} enables users to write mappers that minimize data communication, achieving up to \primitivespeedup performance improvement over a commonly used heuristic in the Chapel framework~\cite{chamberlain2007parallelchapel} for handling dimensionality differences.

While inspired by array programming systems such as APL~\cite{iverson1962programming}, ZPL~\cite{lin1993zpl}, and NumPy~\cite{harris2020array}, \name{} extends these ideas to distributed task-based programs. The processor space transformation is central to the \emph{decompose} primitive, which has no counterpart in prior work. Moreover, \name{} provides a richer set of mapping decisions, including processor selection, memory placement, data layout, garbage collection, and load balancing (see \Cref{sec:discussion}). These capabilities enable \name{} to support a broader range of distributed applications, including tensor algebra and scientific simulations, going beyond prior distributed scheduling systems such as Distributed Halide~\citep{denniston2016distributed}, which are restricted to stencil computations and CPU-only environments.

While \name{}’s transformation primitives resemble classical loop transformations such as tiling, fusion, and fission~\cite{wolf1992improving,chen2008framework,ragan2013halide,chen2018tvm}, they differ in purpose and scope. Loop transformations schedule loops, whereas \name{} schedules sets of tasks across distributed machines. Likewise, \name{} shares the goal of separating performance decisions from algorithms with prior scheduling languages~\cite{zhang2018graphit,senanayake2020sparse,ahrens2022autoscheduling}, but these do not handle dimensionality mismatches between index spaces or explicitly minimize inter-processor data movement.

We implement \name{} on top of the Legion~\cite{bauer2012legion} task-based runtime by translating \name{} mappers to Legion's low-level C++ mapping interface. To evaluate \name{}, we apply it to a set of nine applications, including six advanced distributed matrix multiplication algorithms and three scientific computing workloads. \name{} reduces the lines of code required to implement these mappers by \locreduction{} compared to hand-written C++ mappers, without introducing any observable performance overhead compared to the same mapper written to the C++ interface. In addition, we demonstrate that \name{} enables effective performance tuning: mappers written in \name{} outperform expert-crafted C++ mappers by up to \mapperspeedup{}.

In summary, the contributions of this work are:
\begin{enumerate}
\item A high-level programming interface for mapping distributed applications.
\item A set of transformation primitives for resolving dimensionality mismatches between task and processor spaces, including a key primitive, \emph{decompose}, which helps minimize data communication volume.
\item An implementation of \name{} that reduces mapper code size by \locreduction{} compared to the low-level interface, without incurring additional overhead.
\item A comprehensive evaluation showing that mappers written in \name{} outperform existing expert-written C++ mappers by up to \mapperspeedup{}, and that the \emph{decompose} primitive yields up to \primitivespeedup performance improvement.
\end{enumerate}

\section{Design Goals and Non-Goals}

\begin{figure*}[!tb]
  \centering

  \begin{subfigure}[t]{0.40\textwidth}
    \lstset{style=dslstyle}
    \begin{lstlisting}
m = Machine(GPU)

def block2d(Tuple point, Tuple space):
    idx = point * m.size / space
    return m[*idx]


IndexTaskMap tiles block2d


Region task_init * GPU FBMEM


Layout task_finish * CPU C_order


GarbageCollect systolic *


Backpressure systolic 1


\end{lstlisting}
    \caption{A \name{} mapper.}
    \label{fig:dslmapper}
  \end{subfigure}
  \hfill
  \begin{subfigure}[t]{0.58\textwidth}
    \lstset{style=cppstyle}
    \begin{lstlisting}
ShardID shard(const DomainPoint& point,
              const Domain& space,
              const size_t num_nodes) {
  auto rect2d = Rect<2>(space);
  (*@\textcolor{red}{// Implementation of external helper function is omitted}@*)
  auto blocks = this->select_num_blocks<2>(num_nodes, rect2d);
  Point<2> zeroes{0}, ones{1};
  Rect<2> blockSpace(zeroes, blocks - ones);
  auto numPoints = rect2d.hi - rect2d.lo + ones; 
  Point<2> projected = point * blocks / numPoints;
  AffineLinearizedIndexSpace<2> linearizer(blockSpace); 
  return linearizer.linearize(projected);
};

void slice(const Task& task,
         const SliceTaskInput &input,
         SliceTaskOutput &output) {
  vector<Processor> targets = this->select_targets(task);
  auto size = targets.size();
  DomainT<2> space = input.domain;
  Point<2> n_b = this->select_num_blocks<2>(size, space); 
  Point<2> zeros{0}, ones{1};
  (*@\textcolor{red}{... // 125 lines of C++ code omitted here}@*)
  for (PointInRectIterator<2> it(blocks); it()!=NULL; it++) {
    Point<2> lo = *it; Point<2> hi = *it + ones;
    Point<2> slice_lo = n_p * lo / n_b + space.lo;
    Point<2> slice_hi = n_p * hi / n_b + space.lo - ones;
    DomainT<2,coord_t> slice_space; TaskSlice slice;
    slice.domain = {slice_lo, slice_hi};
    slice.proc = targets[index++ % targets.size()];
    output.slices.push_back(slice);
  }
}
\end{lstlisting}
    \caption{Simplified code excerpts from a C++ mapper.}
    \label{fig:cppmapper}
  \end{subfigure}

  \begin{tikzpicture}[overlay, remember picture]
    \definecolor{mapperblue}{RGB}{35,110,230}
    \filldraw[fill=mapperblue!20, draw=mapperblue, opacity=0.25]
      (-7,10.1) rectangle (-2.05,7.7);
    \filldraw[fill=mapperblue!20, draw=mapperblue, opacity=0.25]
      (-1.2,10.1) rectangle (6.8,6.3);     
    \filldraw[fill=mapperblue!20, draw=mapperblue, opacity=0.25]
      (-1.2,6.1) rectangle (6.8,0.75);    
    \draw[->, thick, mapperblue, opacity=0.9]
    (-2.05,9.5) -- (-1.3,9.5);   

    \draw[->, thick, mapperblue, opacity=0.9]
    (-2.05,7.7) -- (-1.3,4.5);   
  \end{tikzpicture}

\caption{Comparison between a \name{} mapper and its partial C++ counterpart. The \name{} mapper uses a high-level, declarative design that abstracts away the complexity of low-level C++ implementations while still supporting performance optimization. The boxed \CodeIn{block2d} function is realized through two separate APIs in the C++ mapper, illustrating the conciseness of \name{}.}
\label{fig:mappercompare}
\end{figure*}

Our goal is to provide a high-level programming interface that abstracts away the complexities of the underlying runtime, allowing users to focus on mapping logic rather than system internals. Specifically, \name{} enables users to place tasks across distributed machines, determine where data resides in memory,  control data layout, manage garbage collection, and make scheduling decisions. We introduce abstractions and transformation primitives that make such decisions directly programmable and easy to change. For example, \CodeIn{block2d} specifies task space distribution, \CodeIn{Region} and \CodeIn{Layout} control data placement and layout, \CodeIn{GarbageCollect} handles memory management, and \CodeIn{Backpressure} guides scheduling. The \name{} mapper in \Cref{fig:mappercompare} achieves the same functionality as the C++ implementation from prior work~\cite{yadav2022distal} but with significantly less code and greater clarity. We quantify the reduction in code size in \Cref{sec:result:locreduction}.

\Cref{fig:dslmapper} presents only a subset of \name{}'s features. We present the complete feature set of \name{} in \Cref{sec:discussion}. The effectiveness of \name{} for performance tuning is demonstrated in \Cref{sec:result:performance}. We provide a full \name{} mapper in the supplementary material (\Cref{subsec:cannon}).

We do not aim to support portability across different runtime systems. This non-goal reflects the inherent differences among distributed runtimes, which often prevent direct reuse of implementations. Nonetheless, we believe the ideas and abstractions introduced here can inform the design of mapping interfaces for other distributed systems.

\section{A Core Challenge: Mapping Task Spaces to Processor Spaces}
\label{sec:method}

A key challenge for any mapping interface in distributed applications is assigning tasks to processors. The algorithm defines a multi-dimensional \emph{task space} consisting of indexed tasks, while the hardware defines a multi-dimensional \emph{processor space}, capturing its hierarchical structure (e.g., a cluster of $n$ nodes with $m$ GPUs per node). The core problem, \emph{index mapping}, is to define a function mapping task indices in the task space to processor indices in the processor space that minimizes data movement, which is crucial for high performance on distributed systems. Differences in dimensionality between the two spaces further complicate this task.

Task-based systems vary in how they define task spaces.  Some do it implicitly (e.g., through iteration spaces of loops) and some do it explicitly (e.g., Legion's index tasks).   We assume we are given one or more task spaces to map.

We first survey existing approaches to index mapping. Motivated by their limitations, we introduce \name{}'s transformation primitives designed for index mapping in distributed systems. We defer the discussion of how \name{} handles dimensionality mismatches between task and processor spaces, using the \emph{decompose} primitive, to~\Cref{sec:decompose}.

\begin{figure}
    \centering\includegraphics[keepaspectratio=true,width=0.92\columnwidth]{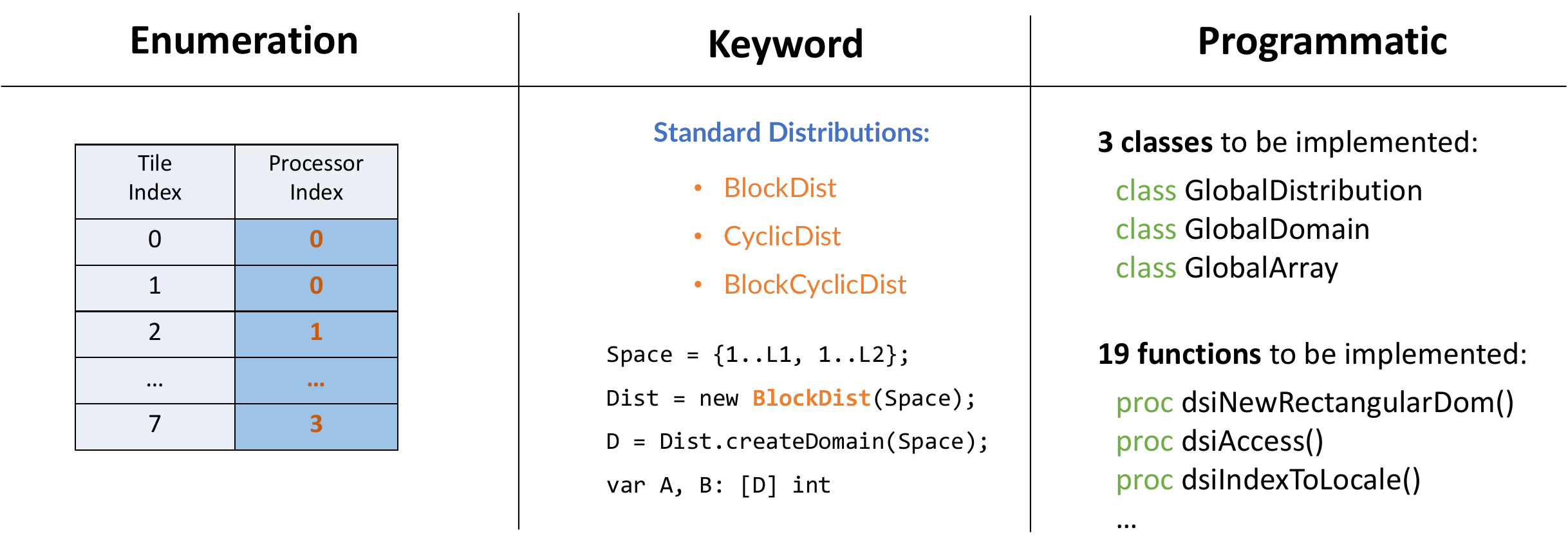}
    \precaptionspace
\caption{Three existing interface designs for mapping task space to processor space: the enumeration-based, keyword-based, and programmatic approaches.}
\label{fig:threedesigns}
\end{figure}

\subsection{Existing Approaches}
\label{subsec:existing}

Existing task-based systems support mapper development through one of three design styles, as illustrated in \Cref{fig:threedesigns}. The \emph{enumeration-based} interface~\cite{ampi} allows users to explicitly specify how each tile of the distributed array is mapped to a processor. While this interface is expressive, it lacks generality: enumeration-based mappers are hard-coded to a specific size of inputs or machines.

In contrast, \emph{keyword-based} designs~\cite{keyword} provide a higher-level abstraction by letting users choose from a predefined set of standard distributions (e.g., \CodeIn{BlockDist} for block distribution). However, this approach sacrifices flexibility—the fixed keyword-based options cannot express non-standard mapping strategies. For example, none of the distributions used in the matrix-multiplication algorithms in \Cref{sec:result:locreduction} can be represented using keyword-based interfaces.

The \emph{programmatic} approach defines an interface of classes and methods that must be implemented to control mapping behavior. While this design offers flexibility, existing programmatic interfaces expose significant system detail and complexity. For instance, defining a new mapping in Chapel requires implementing 19 functions based on 33 pages of documentation~\cite{chamberlain2010user,chamberlain2011authoring}, and Legion's programmatic interface directly exposes the runtime's internal abstractions, accompanied by 19 pages of documentation~\cite{bauer2012legion}. In our experience, only users with substantial expertise in the underlying runtime system are able to successfully develop mappers using such interfaces.

\subsection{Motivating Examples}
\label{subsec:motivating}

\begin{figure}
    \centering\includegraphics[keepaspectratio=true,width=0.9\columnwidth]{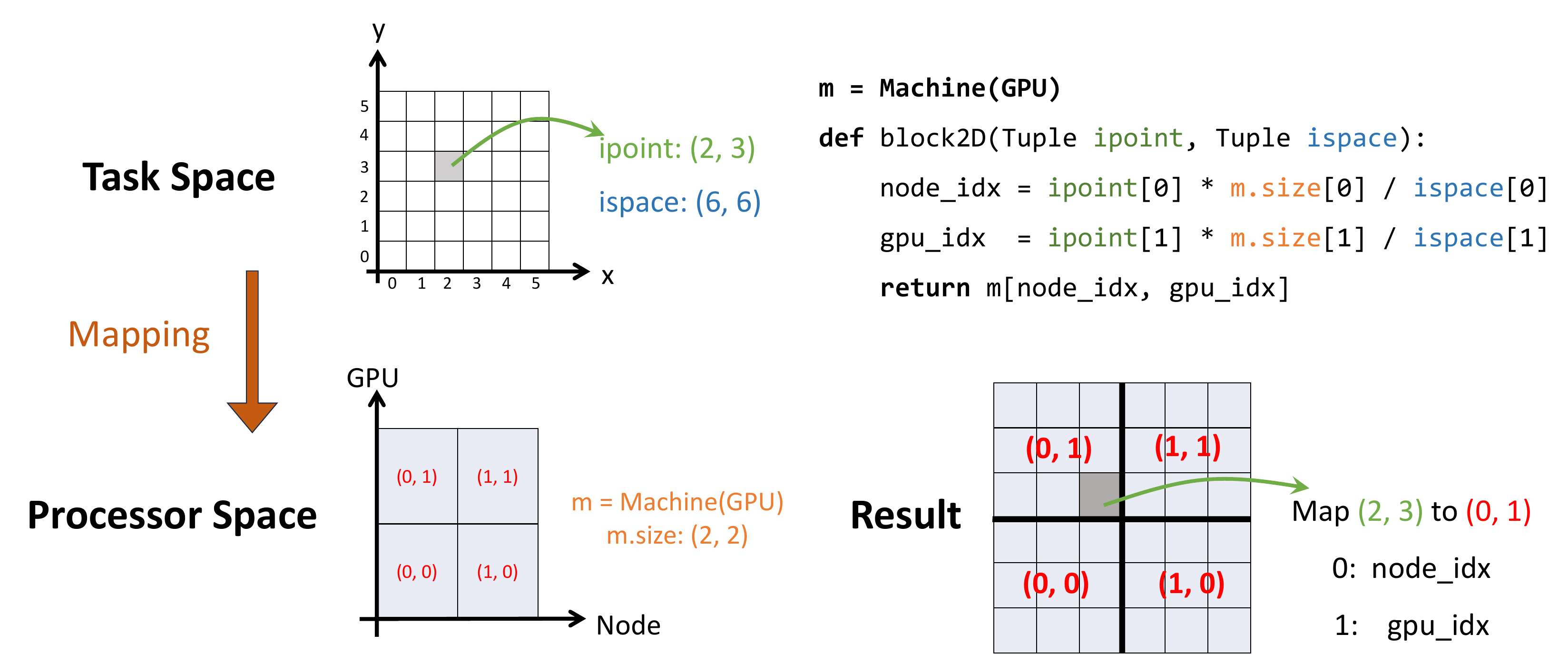}
    \precaptionspace
    \caption{Block mapping from the task space  $(6, 6)$ to the processor space  $(2, 2)$,  a machine with 2 nodes and 2 GPUs per node. A node index and a GPU index within the node name a specific GPU processor. The shaded index point $(2, 3)$ is mapped to node 0 and GPU 1.}
    \label{fig:twoDblock}
\end{figure}

We present three example mappers in \name{} to illustrate how task spaces are mapped onto processor spaces. The task space comes from the application code, which is separate from the mapper. We show how tasks and task spaces are defined in \Cref{subsec:cannon} of the supplementary material.

Below, we describe a standard block distribution in \Cref{subsubsec:standardblock} to introduce the core concepts. Next, we discuss a custom cyclic distribution in \Cref{subsubsec:nonstandard}, which is not supported in keyword-based mapping interfaces. Finally, we use a mapper from Solomonik's 2.5D matrix multiplication algorithm~\cite{solomonik2011communication} in \Cref{subsubsec:mismatch} to demonstrate the complexity introduced by mismatched dimensionality between the task and processor spaces.

\subsubsection{Standard Block Distribution}
\label{subsubsec:standardblock}

Block distribution is a common default strategy widely supported by parallel programming systems. We use it here to illustrate the core concepts of index mapping in \name{}. In \name{}, users define mappers by writing a function that maps each point in the task space to a target processor in the processor space. \Cref{fig:twoDblock} shows the function \CodeIn{block2D}, which implements a block distribution for a task space of size $(6,6)$ (that is, $0 \leq x < 6$ and $0 \leq y < 6$) and a processor space of size $(2,2)$, representing two nodes with two GPUs per node. The function assigns each point to a specific processor.

In \name{}, the GPU processor space can be accessed by \CodeIn{Machine(GPU)}, which is a 2D tuple whose shape, representing the number of nodes and GPUs per node, is automatically inferred by the runtime. Consequently, users do not need to specify these dimensions explicitly, allowing \name{} mappers to remain portable across machines of different sizes. In the mapper function, the index point \CodeIn{ipoint}, task space \CodeIn{ispace}, and processor space size \CodeIn{m.size} are all 2D integer tuples. These values are used to compute the processor indices \CodeIn{node\_idx} and \CodeIn{gpu\_idx}, which specify the destination processor for each index point. As shown in \Cref{fig:twoDblock}, the shaded point $(2, 3)$ is mapped to node 0 and GPU 1. The function returns a specific processor for each index point, resulting in a block distribution over the processor space.

\begin{figure}
    \centering\includegraphics[keepaspectratio=true,width=0.9\columnwidth]{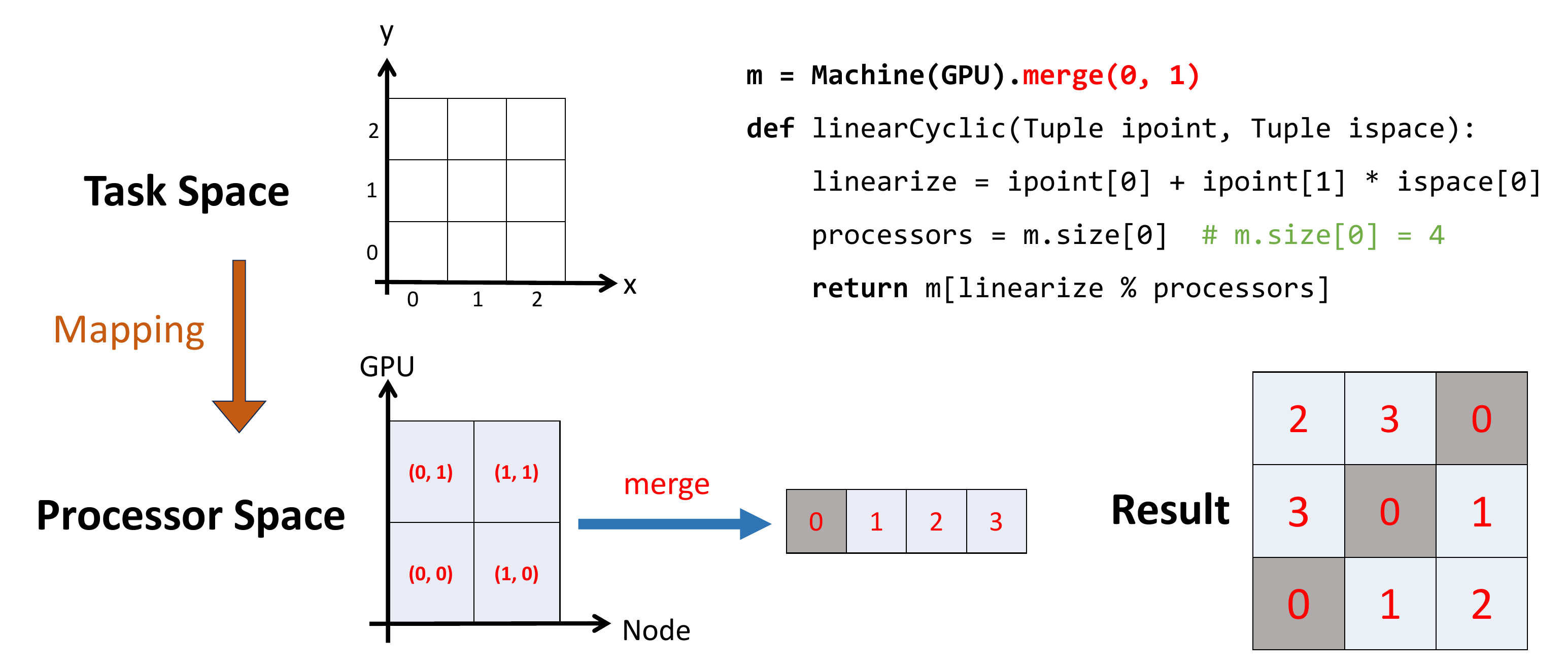}
    \precaptionspace
    \caption{A custom cyclic distribution. The \CodeIn{merge} primitive transforms the 2D processor space into a 1D space. The mapping function linearizes each 2D index point and applies a round-robin distribution over the resulting 1D processor space.}
    \label{fig:nonstandard}
\end{figure}

\subsubsection{Custom Cyclic Distribution}
\label{subsubsec:nonstandard}

We present a custom distribution expressible in \name{} but unsupported by most parallel programming systems that limit users to predefined keyword-based schemes. This non-standard cyclic distribution, inspired by a simplified variant of a matrix multiplication algorithm, is implemented by the mapping function \CodeIn{linearCyclic} in \Cref{fig:nonstandard}, where the subdiagonal index points (shaded) map to the first processor.

The mapper first applies the \emph{merge} primitive, which fuses dimensions 0 and 1 of the 2D processor space into a single dimension (formally defined in \Cref{subsec:transformation}). The resulting processor space \CodeIn{m} is one-dimensional with size 4. The 2D task index points are then linearized and assigned to processors in a round-robin fashion.

\subsubsection{Mapping Under Dimensionality Mismatch}
\label{subsubsec:mismatch}

\begin{figure}
    \centering\includegraphics[keepaspectratio=true,width=\columnwidth]{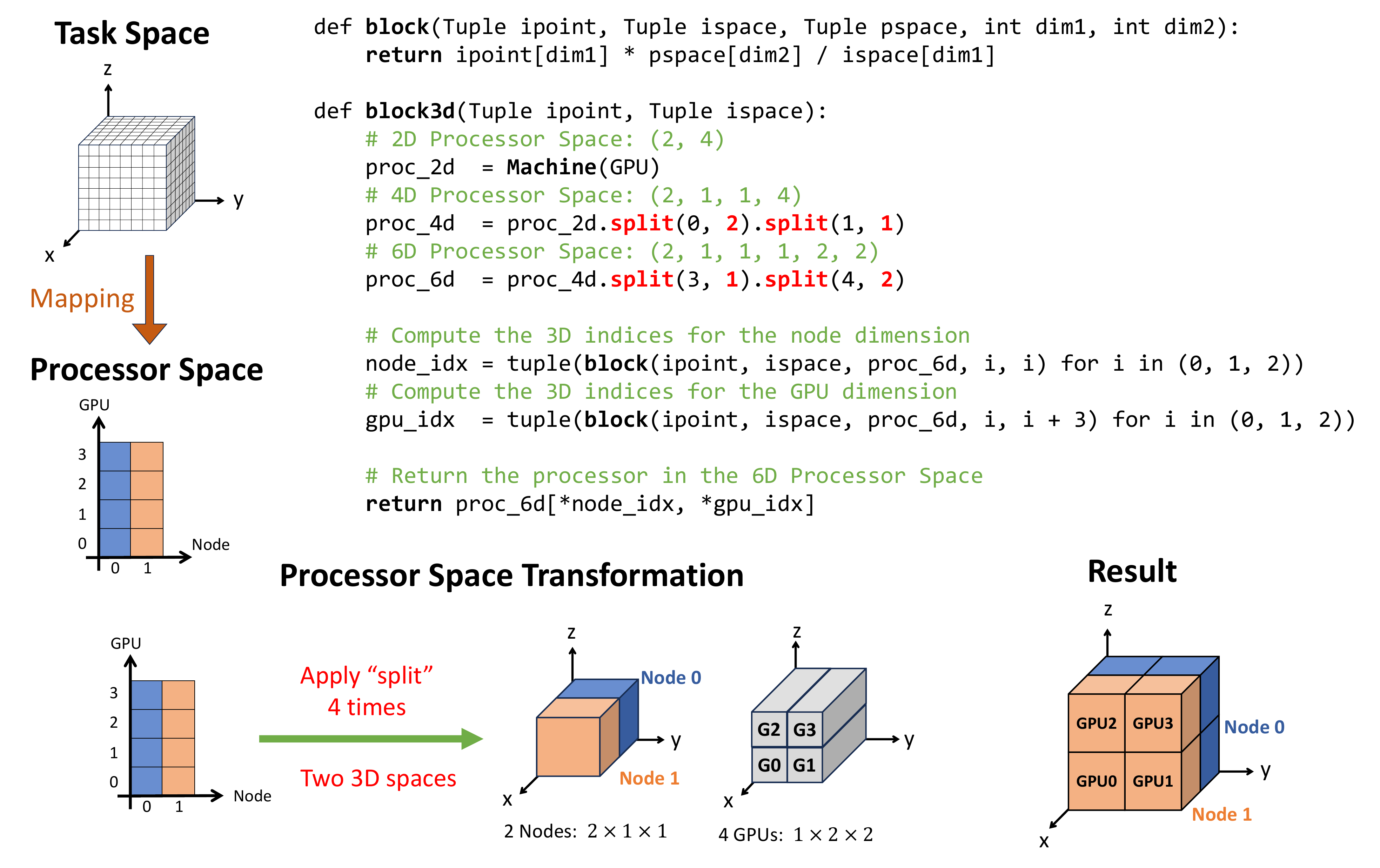}
    \precaptionspace
    \caption{A mapper illustrating the dimensionality mismatch between task and processor spaces in the \solomonik algorithm on a 2-node, 4-GPU-per-node machine. The original 2D processor space is transformed to 6D via the \emph{split} primitive, shown as two 3D spaces.}
    \label{fig:solomonik}
\end{figure}

We give an example illustrating the complexity introduced by a mismatch between the dimensionality of the task space and the processor space. \Cref{fig:solomonik} shows a mapper for the \solomonik algorithm executed on a 2-node machine, where each node has 4 GPUs. The task space is three-dimensional, and the algorithm specifies a hierarchical distribution: the task space is first partitioned along the x-axis so that each node handles half of the x-dimension, and within each node, the 4 GPUs perform a 2D block distribution over the y-z plane. 

The task space is three-dimensional, while the processor space is initially two-dimensional. To support the mapping required by the algorithm, we apply the \emph{split} primitive four times (marked in red in the code). The first two splits align the node dimension with the task space’s x-axis, and the last two align the GPU dimensions with the y- and z-axes. This produces a six-dimensional processor space, visualized as two 3D spaces: one for nodes $(2, 1, 1)$ and one for GPUs $(1, 2, 2)$. The \emph{split} primitive is formally defined in \Cref{subsec:transformation}.

This example also raises deeper research questions. Can we automatically determine the appropriate splitting factors? What underlying principles guide the mapping decisions when resolving mismatched dimensions? Is it possible to address dimensionality mismatches through a generalized transformation? We explore these questions in \Cref{sec:decompose}, where we introduce a new primitive called \emph{decompose} to provide a unified solution.

\subsection{Transformation Primitives}
\label{subsec:transformation}

\begin{figure}
\centering
\scalebox{0.9}{
\begingroup
\renewcommand{\arraystretch}{1.2} 
\begin{tabular}{c|p{7.5cm}}
\hline
\textbf{Transformation} & \parbox[c]{7.5cm}{\centering \textbf{Semantics}} \\
\hline
\multirow{4}{*}{$m' = m.\mathtt{split}(i, d)$} & $ m'[a_0, \ldots, a_n] \coloneq m[b_0, \ldots, b_{n-1}]$\\
& $b_t = \begin{cases}
   a_t \qquad & \quad t < i\\
   a_i + a_{i+1} \cdot d \qquad & \quad t = i\\
   a_{t+1} \qquad & \quad t > i
\end{cases}$\\
\hline
\multirow{5}{*}{$m' = m.\mathtt{merge}(p, q)$} & $ m'[a_0, \ldots, a_{n-2}] \coloneqq m[b_0, \ldots, b_{n-1}]$\\
 & 
$b_t = \begin{cases}
   a_t \qquad & t < p \text{ or } p < t < q\\
   a_p \mathbin{\bmod} m.\mathtt{size}[p] \qquad & t = p\\
   \left\lfloor \frac{a_p}{m.\mathtt{size}[p]} \right\rfloor \qquad & t = q\\
   a_{t-1} \qquad & t > q\\
\end{cases}$\\
\hline
\multirow{4}{*}{$m' = m.\mathtt{swap}(p, q)$} & $m'[a_0, \ldots, a_{n-1}] \coloneq m[b_0, \ldots, b_{n-1}]$\\
& $b_t = \begin{cases}
   a_q              \qquad & \quad t = p\\
   a_p              \qquad & \quad t = q\\
   a_t              \qquad & \quad t \neq p \land t\neq q\\
\end{cases}$\\
\hline
\multirow{2}{*}{$m' = m.\mathtt{decompose}(i, T)$} & {$T = (l_1, \ldots, l_k)$}\\
 & Formally defined in \Cref{sec:decompose}\\
\hline
\end{tabular}
\endgroup
}
\caption{Semantics of transformation primitives expressed as mappings from the indices of the transformed processor space to the indices of the original processor space.}
\label{fig:transprimitive}
\end{figure}

We define the semantics of each \name{} transformation primitive in \Cref{fig:transprimitive}, with the exception of the \emph{decompose} primitive, which is deferred to \Cref{sec:decompose}. Each transformation primitive is a function that takes a processor space~$m$ as input and returns a transformed processor space~$m'$, following the mapping illustrated on the right-hand side of \Cref{fig:transprimitive}. We now describe each transformation.

The {\em split} transformation takes two arguments: a dimension index~$i$ and a splitting factor~$d$. Given a processor space~$m$ of shape $(s_0, \dots, s_{n-1})$, the operation $m' = m.\mathtt{split}(i, d)$ produces a new processor space~$m'$ of shape $(s_0, \dots, d, s_i / d, \dots, s_{n-1})$. Like other processor space transformations, this transformation is invertible; in our implementation, \name{} uses the inverse to translate from mapper operations on $m'$ to the original space $m$. 

The {\em merge} transformation takes two dimensions~$p$ and~$q$ of the processor space and fuses them into one. Given a processor space~$m$ of shape $(s_0, \dots, s_{n-1})$, the operation $m' = m.\mathtt{merge}(p, q)$ produces a new processor space~$m'$ of shape $(s_0, \dots, s_p \cdot s_q, \dots, s_{n-1})$, where the two dimensions at positions~$p$ and~$q$ in~$m$ are combined into a single dimension at position~$p$ in~$m'$. The index mapping is given by: $m'[a_0, \dots, a_{n-2}] = m[a_0, \dots, a_p \bmod s_p, \dots, \lfloor a_p / s_p \rfloor, \dots, a_{n-2}]$.

Transformation primitives in \name{} can be composed sequentially. Consider a 2D processor space~$m$, and let $m' = m.\mathtt{split}(0, d)$, followed by $m'' = m'.\mathtt{merge}(0, 1)$. The resulting processor space~$m''$ is again 2D. We now derive the index transformation from~$m''$ to the original space~$m$ by sequentially applying the semantics of the \emph{merge} and \emph{split} transformations: $m''[a_0, a_1] = m'[a_0 \mathbin{\bmod} d, \lfloor a_0 / d \rfloor, a_1] = m[(a_0 \mathbin{\bmod} d) + \lfloor a_0 / d \rfloor \times d, a_1]$. Because the expression $(a_0 \mathbin{\bmod} d) + \lfloor a_0 / d \rfloor \times d$ simplifies to $a_0$ for all integers, it follows that $m''[a_0, a_1] = m[a_0, a_1]$ ---  \emph{merge} is the inverse of \emph{split}.

The {\em swap} transformation exchanges two dimensions of the processor space. Combined with {\em merge}, which flattens two dimensions into one, users can control whether merging follows row-major or column-major order.

\begin{figure}
    \centering\includegraphics[keepaspectratio=true,width=\columnwidth]{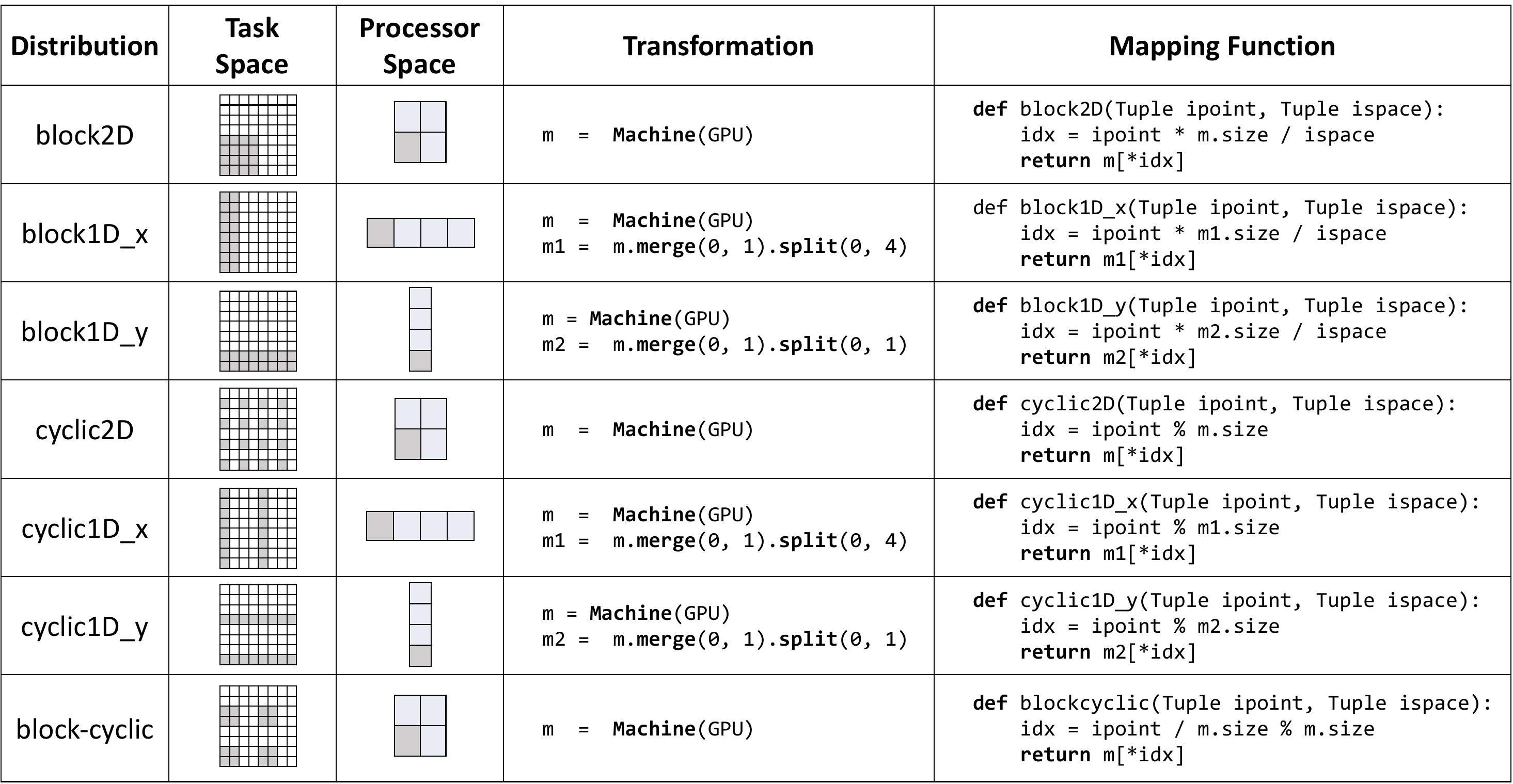}
    \precaptionspace
    \caption{Common distributions expressed in \name{}. The shaded region in the task space is mapped to the corresponding shaded processor. The transformation code reshapes the original $(2, 2)$ processor space into the desired processor space, which is then used in the user-defined mapping function.}
    \label{fig:codecommondist}
\end{figure}

Using these primitives, \name{} can transform processor spaces in flexible ways. \Cref{fig:codecommondist} illustrates how \name{} supports common distribution patterns. Suppose we want to map a 2D task space onto a 2-node machine with 2 GPUs per node. In the figure, the shaded region of the task space is mapped to the corresponding shaded processor in the processor space.

For the \CodeIn{block2D} distribution, the mapping function is equivalent to the one in \Cref{fig:twoDblock}, but is simpler thanks to \name{}’s support for tuple arithmetic. Tuple operands must have the same shape. In \Cref{fig:codecommondist}, both the transformed processor spaces and the task space are 2D. In particular, the \CodeIn{*} operator flattens tuples for indexing: for example, \verb|m[*(1,2)]| is equivalent to \verb|m[1,2]|. \name{} supports both elementwise arithmetic on individual tuple dimensions and operations over entire tuples, enabling users to express index mappings more concisely and intuitively.

\section{Decompose Transformation Primitive}
\label{sec:decompose}

As discussed in \Cref{subsubsec:mismatch}, mismatches between the dimensionality of the task and processor spaces pose significant challenges for mapping decisions. In this section, we address this issue by analyzing its underlying principles and introducing a generalized transformation primitive, \decompose. \Cref{subsec:decompose:comparison} analyzes a suboptimal, but standard, heuristic to handle dimensionality mismatches. We then formally define the \decompose primitive in \Cref{subsec:decompose:def}, grounded in a communication volume analysis. Finally, \Cref{subsec:decompose:impl} describes our search-based optimization algorithm and analyzes its complexity.

\subsection{Suboptimal Existing Heuristics for Resolving Dimensionality Mismatch}
\label{subsec:decompose:comparison}

We examine how existing task-based systems address dimensionality mismatches between task and processor spaces. Most frameworks sidestep this issue by linearizing both spaces and applying a default 1D block mapping. However, they rarely document their linearization procedures, making direct comparison difficult. An exception is Chapel~\cite{chamberlain2007parallelchapel}, a widely used parallel programming framework, which explicitly describes its approach. To expose the limitations of this strategy, we present an example showing how it can lead to suboptimal mappings.

Suppose we have 6 processors and a 2D task space. To match the dimensionality, we must split the 6 processors into a 2D tuple. There are four possible factorizations: $(6, 1)$, $(3, 2)$, $(2, 3)$, and $(1, 6)$. The existing heuristic used to determine the processor grid, as shown in \Cref{alg:chapel} of the supplementary material, does not consider the shape of the task space.

\begin{figure}[!tb]
    \centering\includegraphics[keepaspectratio=true,width=0.9\columnwidth]{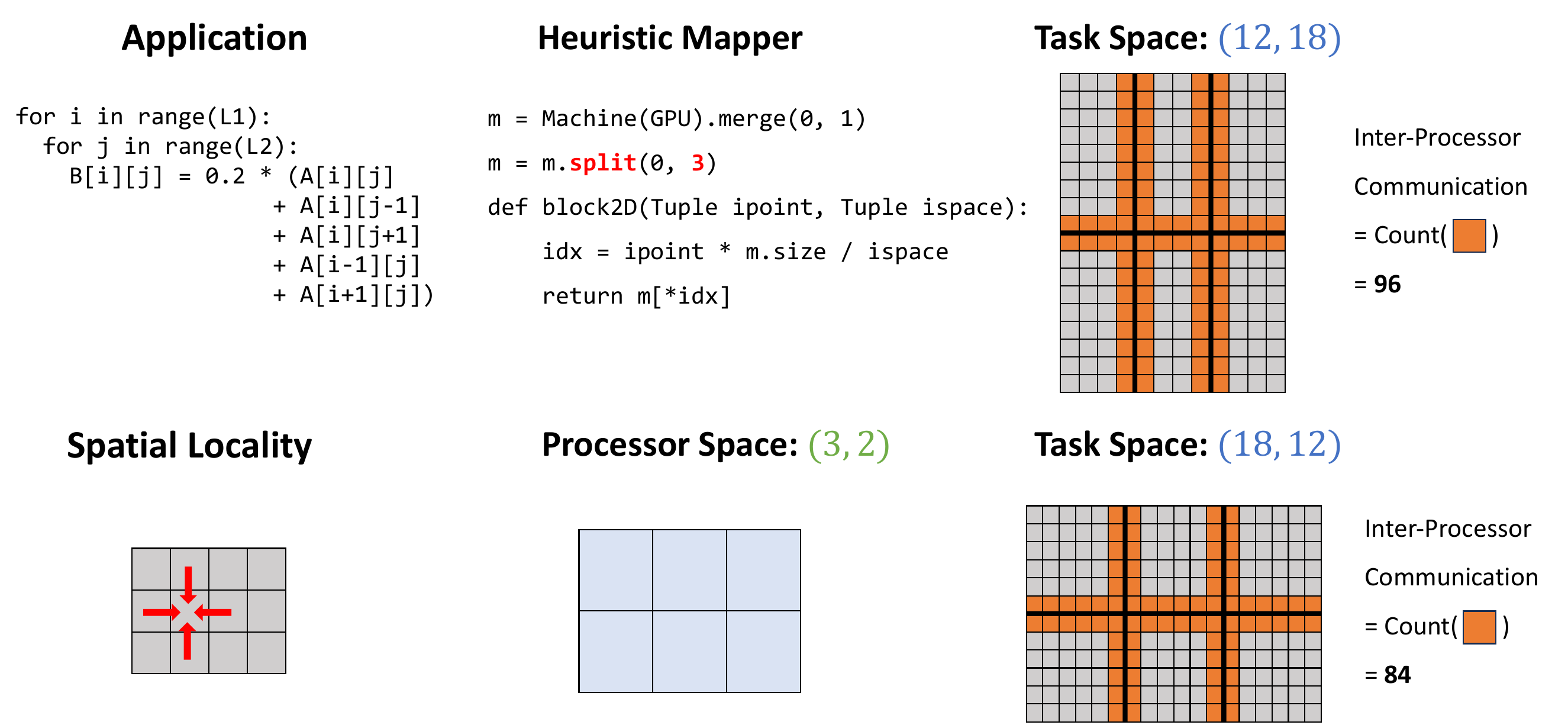}
    \precaptionspace
    \caption{For this application, the mapper from \Cref{alg:chapel} (see supplementary material) selects a fixed $(3,2)$ processor grid for 2D block mapping. Orange regions indicate inter-processor data transfers. The $(12,18)$ task space incurs higher communication than $(18,12)$, showing that the chosen grid is suboptimal; a $(2,3)$ grid would yield lower communication.}
    \label{fig:inputagnostic}
\end{figure}

The heuristic follows a greedy strategy, aiming to produce factors that are as balanced as possible in magnitude. In the example above, the algorithm selects the grid $(3, 2)$ by prioritizing balanced factorization.  Now consider two task spaces, $(12, 18)$ and $(18, 12)$, for an application with nearest neighbor locality (i.e., data is exchanged among immediate neighbors in task space), as shown in \Cref{fig:inputagnostic}. Each task space is mapped to the processor grid $(3, 2)$ using the \texttt{block2D} function, which partitions the task space into rectangular blocks assigned to each processor. The orange regions in the figure indicate data that must be transferred across processor boundaries due to this mapping.

For the $(12, 18)$ task space, the total inter-processor communication volume is 96 elements, while for the $(18, 12)$ task space it is 84 elements. This difference highlights the sensitivity of communication cost to how the task space is aligned with the processor grid. If the algorithm had taken the shape of the task space into account and chosen the processor grid $(2, 3)$ for the $(12, 18)$ space, the communication volume could have matched the more efficient 84-element case, avoiding unnecessary communication overhead.

\subsection{Formal Definition of Decompose and Intuitive Solution}
\label{subsec:decompose:def}

We formally define the optimization problem that the \texttt{decompose} primitive aims to solve, based on an analysis of inter-processor communication volume. We also present the underlying intuition to clarify how the solution addresses this problem.  Implicit in all solutions to this problem (including the greedy heuristic described above) are two assumptions: that tasks in the task space have uniform cost (i.e., are load balanced) and that there is locality in the communication pattern, so that partitioning the task space into compact regions is desirable to minimize communication.

Suppose we apply \texttt{decompose} to a processor space $M$ of size $(d_0, \ldots, d_{n-1})$, written as $M' = M.\texttt{decompose}(i,\allowbreak (l_1, \ldots, l_k))$. The \texttt{decompose} primitive splits the $i$-th dimension $d_i$ into $k$ natural numbers $d_{i_1}, \ldots, d_{i_k}$, producing a new processor space $M'$ of size $(d_0, \ldots, d_{i_1}, \allowbreak \ldots, \allowbreak d_{i_k}, \allowbreak \ldots, d_{n-1})$. The transformation preserves the total size of the processor space, so it must satisfy $\prod_{m = 1}^{k} d_{i_m} = d_i$.

Conceptually, \texttt{decompose} is a shorthand for applying a sequence of \texttt{split} transformations. Specifically, applying $m_k = m_1.\texttt{decompose}(i, T)$ is equivalent to performing $m_{n+1} = m_n.\texttt{split}(i + n - 1, d_{i_n})$ for $1 \leq n < k$. There can be multiple ways to factor $d_i$ into $k$ natural numbers $d_{i_1}, \ldots, d_{i_k}$. The \texttt{decompose} primitive selects a factorization by solving the following optimization problem:

\[
\min_{\{d_{i_m}\in\mathbb{N}\}} \sum_{m=1}^k \frac{d_{i_m}}{l_m}
\quad\text{s.t.}\quad
\prod_{m=1}^k d_{i_m}=d_i.
\]

An equivalent formulation introduces the \emph{workload vector} $w_m = \frac{l_m}{d_{i_m}}$ for $1 \leq m \leq k$, where $w_m$ represents the portion of the task space assigned to each processor in the $m$-th dimension. Rewriting the problem in terms of the workload vector gives:

\[
\min_{\{w_m\}} \sum_{m=1}^k \frac{1}{w_m}
\quad\text{s.t.}\quad
\prod_{m=1}^k w_m = \frac{\prod_{m=1}^k l_m}{d_i},\;
\frac{l_m}{w_m} \in \mathbb{N}.
\]

\begin{figure}
    \centering
    \includegraphics[keepaspectratio=true,width=0.85\columnwidth]{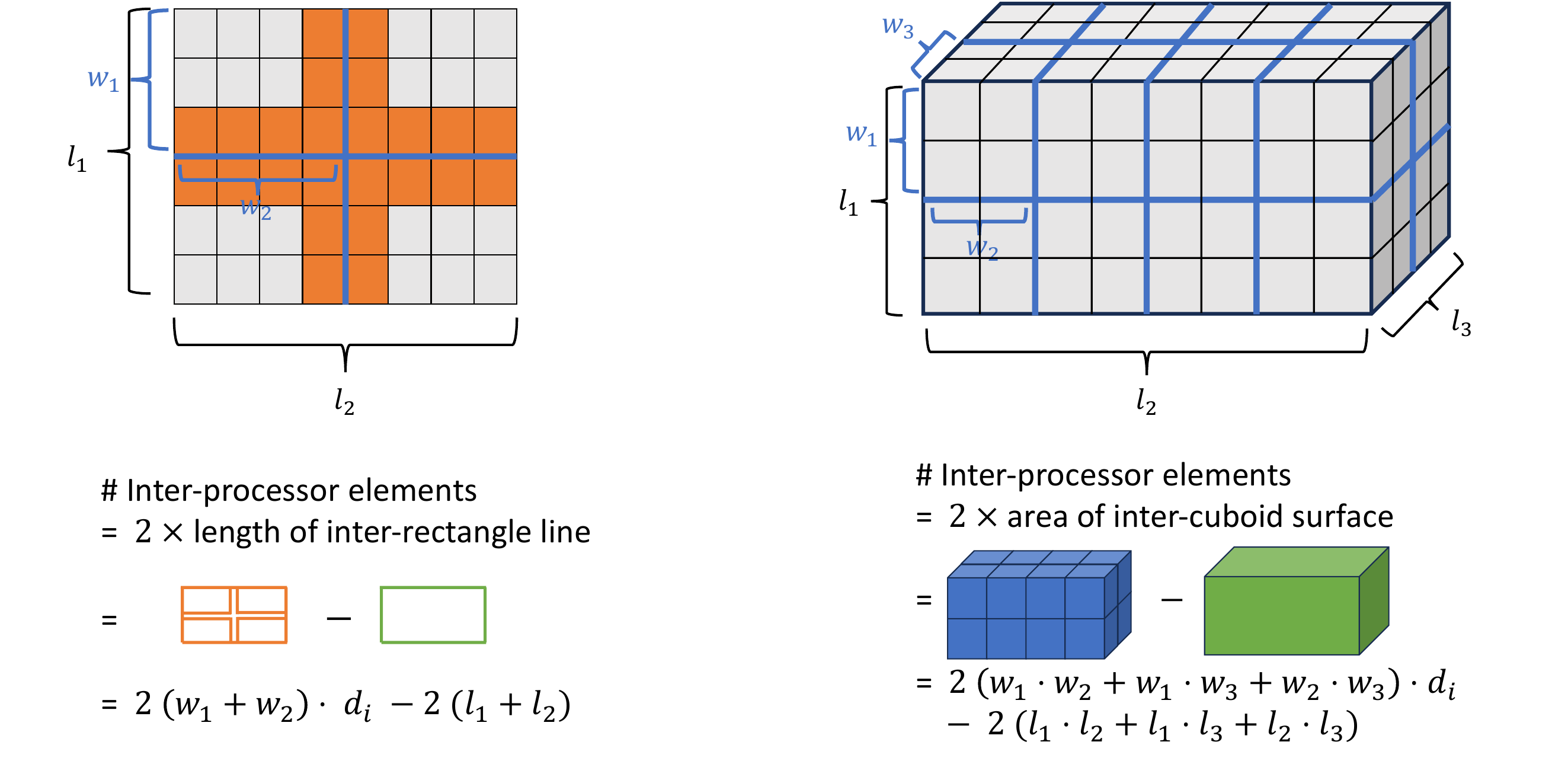}
    \precaptionspace
    \caption{We show two examples of how to compute the communication volume (namely, the number of inter-processor elements). In the 2D example, we compute the length of the inter-rectangle line by summing up the perimeter of all 4 rectangles of $(w_{1}, w_{2})$ and subtracting the perimeter of the $(l_{1}, l_{2})$ rectangle. In the 3D example, we compute the area of the inter-cuboid surface by summing up the surface area of all 16 cuboids of $(w_{1}, w_{2}, w_{3})$ and subtracting the surface area of the $(l_{1}, l_{2}, l_{3})$ cuboid.}
    \label{fig:formula}
\end{figure}

We now illustrate the optimization problem using the two examples shown in \Cref{fig:formula}. The left-hand side depicts a 2D case with $k = 2$ and processor count $d_i = 4$. To compute the communication volume, we count the number of elements transferred across processor boundaries. This volume equals $2 \cdot (w_1 + w_2) \cdot d_i - 2 \cdot (l_1 + l_2)$, where the first term is the total perimeter of the $d_i$ blocks of size $(w_1, w_2)$, and the second term is the perimeter of the entire task space of size $(l_1, l_2)$. 

Given fixed input dimensions $l_1$ and $l_2$ and processor count $d_i$, the second term is constant, and $w_1 \cdot w_2 = \frac{l_1 \cdot l_2}{d_i}$ is also fixed. Therefore, minimizing the communication volume reduces to minimizing $w_1 + w_2$, which is equivalent to minimizing $\frac{1}{w_1} + \frac{1}{w_2}$. This matches the objective of our optimization formulation in the 2D case.

On the right-hand side, the figure shows a 3D example with $k = 3$ and $d_i = 16$. Here, the communication volume corresponds to the total inter-block surface area. The surface area $S$ of the $d_i$ cuboids of size $(w_1, w_2, w_3)$ is given by $2S = 2(w_1 w_2 + w_1 w_3 + w_2 w_3) \cdot d_i - 2(l_1 l_2 + l_1 l_3 + l_2 l_3)$.
Given fixed input dimensions $l_1, l_2, l_3$ and processor count $d_i$, the term $w_1 w_2 w_3 = \frac{l_1 l_2 l_3}{d_i}$ is constant. Thus, minimizing $S$ is equivalent to minimizing $w_1 w_2 + w_1 w_3 + w_2 w_3$. Under the fixed-product constraint, this can be transformed into minimizing $\frac{1}{w_1} + \frac{1}{w_2} + \frac{1}{w_3}$, again aligning with the objective of our optimization formulation in the 3D case. This analysis generalizes to any $k$-dimensional case. Let $S$ denote the total inter-surface area in the $k$-dimensional setting. Then,
\begin{equation*}
2S = SA(w_1, w_2, \ldots, w_k) \cdot d_i - SA(l_1, l_2, \ldots, l_k),
\label{eqn:comm_area}
\end{equation*}
where $SA(x_1, x_2, \ldots, x_k)$ denotes the surface area of a $k$-dimensional hyperrectangle, defined as
\[
SA(x_1, x_2, \ldots, x_k) = 2 \cdot \left( \prod_{m=1}^{k} x_m \right) \cdot \left( \sum_{m=1}^{k} \frac{1}{x_m} \right).
\]
Given that $\prod_{m=1}^{k} w_m = \frac{\prod_{m=1}^{k} l_m}{d_i}$ is constant for fixed input size and processor count, minimizing $S$ reduces to minimizing $\sum_{m=1}^{k} \frac{1}{w_m}$—which matches the objective of the optimization problem introduced earlier.

The solution to the optimization problem is not immediately obvious. In the following, we present an intuitive justification for the solution, which is based on the classic inequality stated below.

\begin{theorem*}
Given a list of $n$ positive numbers $a_1, a_2, \ldots, a_n$, the following inequality holds:
\[
\frac{1}{n} \sum_{m=1}^{n} a_m \geq \left( \prod_{m=1}^{n} a_m \right)^{1/n},
\]
with equality if and only if $a_1 = a_2 = \cdots = a_n$.
\end{theorem*}

This is the well-known arithmetic-geometric mean (AM-GM) inequality. It states that the arithmetic mean of a set of positive numbers is always greater than or equal to their geometric mean, with equality only when all values are equal. In the optimization problem, the term $\frac{\prod_{m=1}^{k} l_m}{d_i}$ is a constant determined by the input size and the number of processors. We can now apply the inequality to the objective: $\frac{1}{k} \sum_{m=1}^{k} \frac{1}{w_m} \geq \left( \prod_{m=1}^{k} \frac{1}{w_m} \right)^{1/k}$. Given $\prod_{m=1}^{k} w_m = \frac{\prod_{m=1}^{k} l_m}{d_i}$, we obtain $\sum_{m=1}^{k} \frac{1}{w_m} \geq k \cdot \left( \frac{d_i}{\prod_{m=1}^{k} l_m} \right)^{1/k}$. Equality is achieved when all terms $\frac{1}{w_m}$ are equal, which implies $w_1 = w_2 = \cdots = w_k$. In practice, however, exact equality may not be attainable because each $\frac{l_m}{w_m}$ must be a natural number.

In the 2D task space example shown in \Cref{fig:inputagnostic}, the mapping for the $(18, 12)$ space is optimal because the workload vector is $(w_1, w_2) = \left(\frac{l_1}{d_1}, \frac{l_2}{d_2}\right) = (6, 6)$, where the two components are equal. Similarly, the 3D example in \Cref{fig:formula} uses a task space of $(4, 8, 4)$ mapped onto 16 processors. The solution shown in the figure yields a workload vector $(w_1, w_2, w_3) = (2, 2, 2)$, which satisfies the sufficient condition for attaining the minimum communication volume.

These examples suggest that the optimal solution is achieved when the task space is divided among the processor dimensions in a balanced manner. However, in practice, this ideal may not be attainable due to the integrality constraint that each $\frac{l_m}{w_m}$ must be a natural number. We introduce a search-based algorithm for solving the optimization problem in \Cref{subsec:decompose:impl}.

As mentioned above, our analysis assumes locality in communication (e.g., such as nearest-neighbor communication), which is very common, but not universal, in practice. Other typical patterns, including anisotropic halo and all-to-all exchanges, are discussed in \Cref{subsec:generalization} (see supplementary material). For algorithms without spatial locality or highly unbalanced tasks, \emph{decompose} may be ineffective. In such cases, a good \name{} mapper may rely on other features, such as load-balancing (see \Cref{sec:discussion}).

\subsection{Algorithm Implementation and Complexity Analysis}
\label{subsec:decompose:impl}

At a high level, the integrality constraint requires us to enumerate all possible ways of factoring $d_i$ into $k$ positive integers $d_{i_1}, \ldots, d_{i_k}$ such that $\prod_{m=1}^{k} d_{i_m} = d_i$, and then select the factorization that minimizes the objective function $\sum_{m=1}^{k} \frac{d_{i_m}}{l_m}$. The main challenge lies in efficiently and exhaustively enumerating all valid factorizations to ensure optimality.

We begin with a simple case: suppose $k = 3$ and $d_i = 16 = 2^4$. To enumerate all possible factorizations, we must determine all ways to distribute the four factors of 2 across the three dimensions. This reduces to finding all non-negative integer solutions to the equation $x_1 + x_2 + x_3 = 4$, which can be solved using recursion or backtracking.

A more complex example is when $k = 3$ and $d_i = 48 = 2^4 \cdot 3^1$. In this case, we must find all non-negative integer solutions to both $x_1 + x_2 + x_3 = 4$ and $y_1 + y_2 + y_3 = 1$, corresponding to the distribution of the prime factors 2 and 3, respectively. Each valid factorization corresponds to a tuple of the form $(2^{x_1} \cdot 3^{y_1}, 2^{x_2} \cdot 3^{y_2}, 2^{x_3} \cdot 3^{y_3})$ for some choice of $(x_n, y_n)$.

In the general case where $d_i = p_1^{a_1} \cdot p_2^{a_2} \cdot \ldots \cdot p_t^{a_t}$, the prime factorization naturally decomposes the enumeration task. We can independently enumerate the placement strategies for each prime factor by solving $t$ separate integer partition problems of the form $z_1 + \ldots + z_k = a_j$ for each $p_j$, and then compute the Cartesian product of their solutions to generate all valid factorizations.

We now analyze the complexity of the search-based algorithm. Since the algorithm exhaustively enumerates all valid factorization strategies, its complexity is determined by the size of the search space. As shown in the earlier example, enumerating non-negative integer solutions to the equation $x_1 + x_2 + x_3 = 4$ is equivalent to counting integer partitions with repetition, which can be mapped to the problem of selecting two dividers among six gaps. This corresponds to the number of combinations $\binom{6}{2} = 15$. In general, for a processor count with prime factorization $d_i = p_1^{a_1} \cdot p_2^{a_2} \cdots p_t^{a_t}$, the total number of factorization strategies is $\prod_{j=1}^{t} \binom{a_j + k - 1}{k - 1}$. Each term in the product counts the number of ways to distribute the $a_j$ copies of prime $p_j$ across $k$ dimensions. In practice, the exponents $a_j$ are usually small (typically less than 10), and the number of dimensions $k$ is often no greater than 3. As a result, the search space remains small, and the algorithm is computationally efficient in real-world scenarios.

We believe that enumeration is necessary to guarantee an optimal solution. Consider a greedy algorithm that, at each step, assigns prime factors of $d_i = 72$ (i.e., $2, 2, 2, 3, 3$) to dimensions in a way that minimizes the maximum difference between elements of the workload vector. Given a task space of $(l_1, l_2) = (8, 9)$, this greedy strategy produces a suboptimal workload vector of $\left(\frac{4}{3}, \frac{3}{4}\right)$, which results in imbalanced partitioning. In contrast, our search-based algorithm identifies the optimal factorization that yields a perfectly balanced workload vector of $(1, 1)$.

\section{Implementation}
\label{sec:implementation}

We describe the translation of \name{} to the mapping interface of task-based runtime systems. These systems employ a highly asynchronous and pipelined execution model, where tasks progress through multiple stages to maximize throughput. Mapping decisions, which determine where tasks and data are placed, must be made at various points along this pipeline and are exposed via many distinct callback functions, each tied to a specific stage in a task's lifecycle. This results in a fragmented and low-level interface that mirrors the internal structure of the runtime rather than providing an intuitive programming model for developers.

The core challenge is to design a high-level, unified mapping abstraction that can be faithfully translated into this low-level execution model. To illustrate this semantic gap, we first describe the runtime's execution semantics in \Cref{subsec:semantics}, then present our translation strategy in \Cref{subsec:translationimpl}.

\subsection{Execution Semantics of the Runtime System}
\label{subsec:semantics}

Task-based runtime systems maximize throughput through an asynchronous, pipelined execution model, where a task advances through several distinct stages during its lifetime. To illustrate the low-level nature of the mapping interface, we present a much simplified version of the execution semantics of the Legion runtime system \cite{bauer2012legion}, focusing on the portion of the pipeline that determines the processor on which a task runs. Our semantics makes explicit a key source of complexity: the interface is tightly coupled to internal execution stages, where what is conceptually a single decision---choosing a processor for a task---is split across multiple pipeline stages at a low level of abstraction. 
The same low-level complexity exists in frameworks like StarPU~\cite{augonnet2010starpu}, Charm++~\cite{kale1993charm++}, and HPX~\cite{kaiser2014hpx}, which expose mapping through asynchronous callbacks and fragmented APIs. In contrast, \name{} provides a high-level DSL that centralizes this logic while still providing the same flexibility.

\ignore{
Before presenting the semantics, we first describe the key relationships between tasks that govern how and when they execute. Consider the following simple program:

\begin{lstlisting}[style=taskcode]
task f(a, b) {
    g(a);    // writes a
    h(b);    // writes b
    k(a, b); // reads a, b and writes b
}
\end{lstlisting}

The task $\tt f(a,b)$ is the \emph{parent} of child tasks $\tt g(a)$, $\tt h(b)$, and $\tt k(a,b)$. Each child begins only after the parent starts, and the parent completes only after all its children finish. Since tasks execute sequentially, parallelism arises from asynchronous launches. The order in which the parent invokes its children induces a program order $\tt \preceq$. In this example, $\tt g(a) \preceq h(b) \preceq k(a,b)$, meaning $\tt k(a,b)$ has \emph{sibling predecessors} $\tt g(a)$ and $\tt h(b)$.

Task dependencies are represented by the relation $\tt \leq$, indicating that one task may read or write a value produced by another. In this example, $\tt g(a) \leq k(a,b)$ and $\tt h(b) \leq k(a,b)$ capture that $\tt k(a,b)$ may read values written by $\tt g(a)$ and $\tt h(b)$, and therefore cannot begin execution until both complete. However, this does not prevent all progress on $\tt k(a,b)$—in particular, its \emph{mapping} (i.e., assignment to a processor) can proceed before its dependencies finish. Since there is no dependency between $\tt g(a)$ and $\tt h(b)$, they may execute in parallel.
}

\Cref{fig:executionsyntax} defines the structures used in the execution semantics. A task space is a set of multidimensional points, each representing a single point task to be executed.  Working with sets of tasks is important for the efficiency of the runtime system itself, especially in the early stages of the task pipeline, where runtime operations can be done in constant time on a set rather than in linear time on each individual point task in the set.

The execution context consists of a set of tasks $T$\footnote{One major simplification is that we ignore dependencies between tasks; dependencies are not needed to illustrate the pipeline for choosing a task's processor.}, a log of actions on tasks, and two vectors of $n$ queues, one queue per node of the machine.\footnote{We also do not model nodes, using only processors, but in a full description a set of processors is associated with each node.}
When a task space is first enqueued in the runtime system, it goes into a {\em distribution queue}, from where its set of points may be subdivided and sent to other nodes for further processing.  Eventually (perhaps after multiple recursive subdivisions of the set of tasks), a task space is moved to a {\em mapping queue} on a particular processor, where the assignment of the task space to a (potentially different) processor for execution will be made.  This structure, where task spaces are subdivided and spread out to multiple processors for mapping, parallelizes the mapping process, preventing a single processor from becoming a runtime bottleneck. Once a task space is mapped, its point tasks are enqueued for execution on its assigned processor; because our focus is on the mapping process, we do not illustrate task execution here.

\newcommand{\rulename}[1]{\ensuremath{\textsc{#1}}}

\begin{figure*}[t]
 \centering            
\[
\begin{array}{lrl}
\mbox{\em Tasks} & 
\mbox{Point } P & ::= (i_1, \ldots, i_n) \\
& \mbox{Task Space } t & ::= (id, P~\mbox{\bf set}) \\[.2in]
\mbox{\em State} & 
\mbox{Distribution Queue } E & ::= N[n]   \\
&\mbox{Mapping Queue } M & ::= N[n]   \\
&\mbox{Execution Log } L & ::= e~\mbox{\bf set} \\
&\mbox{Execution Log Entry } e & ::= \mbox{enqueued}(t) \mid \mbox{mapped}(t,p) \\
&\mbox{Processor } p & ::= id 
\end{array}
\]
\precaptionspace
\caption{Abstract Syntax for Execution Model}
\label{fig:executionsyntax}

\vspace{0.5em}
  \setlength{\jot}{4pt} 

  \begin{alignat*}{2}   
  &\frac{%
      \begin{array}{l}
        t\in T\\
   \text{enqueued}(t)\notin L\\
      \end{array}}
     {%
       L,E,M\xrightarrow{T}
       L\cup\{\text{enqueued}(t)\},E+_1 t,M}
  &\qquad&\rulename{[Enqueue]}\\[1.5em]
  &\frac{%
      \begin{array}{l}
        \text{SHARD}(t)=\text{distribute}((t_1,p_1),(t_2,p_2))\\
        t=(id,P), \ t_1=(id_1,P_1), \ 
        t_2=(id_2,P_2)\\
        id_1,id_2\text{ fresh (not in $T,E,M,L$)}\\
        P=P_1\cup P_2\\
      \end{array}}
     {L,E,M \xrightarrow{T}
       L, ((E -_p t) +_{p_1}t_1)+_{p_2}t_2,M}
  &\qquad&\rulename{[Distribute]}\\[1.5em]
  &\frac{%
      \begin{array}{l}
        \text{SHARD}(t)=\text{local}(t)\\
        t = (id,P) \\
        E'=E -_i t
      \end{array}}
     {L,E,M \xrightarrow{T}L,E',M+_i t}
  &\qquad&\rulename{[Local]}\\[1.5em]
  &\frac{%
      \begin{array}{l}
        \text{SLICE}(t)=p\\
        M'=M -_i t
      \end{array}}
     {L,E,M \xrightarrow{T}
       L\cup\{\text{mapped}(t,p)\},E,M'}
  &\qquad&\rulename{[Map]}
  \end{alignat*}
  \caption{Formalization of the mapping pipeline.}
  \label{fig:semantics}
\end{figure*}

\Cref{fig:semantics} formalizes the task execution rules described above using an operational semantics. Each judgment has the form  
\[
L,E, M \xrightarrow{T} L', E', M'
\]
which should be read as: given a set of task spaces $T$, the current execution log $L$ and vectors of queues $E, M$,  transition to a new execution log $L'$ and new vectors of queues $E', M'$. The rules rely on two user-supplied mapping functions, {\tt SHARD} and {\tt SLICE}, with the following signatures:
\[
\begin{array}{rl}
\mbox{SHARD: } & t \rightarrow \mbox{distribute}((t \ast p) \ast (t \ast p)) + \mbox{local}(t) \\
\mbox{SLICE: } & t \rightarrow p
\end{array}
\]

The {\tt SHARD} function takes a task space and either partitions it into smaller task spaces for distribution across nodes or assigns it to the local node. {\tt SLICE} then selects a specific processor within the target node for execution. Although these are two \emph{separate} callback functions in the low-level interface, \name{} \emph{unifies} them by interpreting both as part of a single index transformation from task space to processor space.

\Cref{fig:semantics} formalizes the rules described above as execution judgments. The rules use an operator $+_i$, which adds a task space to the $i$th queue
of a vector of queues:
\[ (Q_1,\ldots,Q_i,\ldots Q_n) +_i t = (Q_1,\ldots, t : Q_i ,\ldots Q_n)\]
We also need the dual dequeuing operation:
\[ (Q_1,\ldots,Q_i : t ,\ldots Q_n) -_i t = (Q_1,\ldots, Q_i ,\ldots Q_n) \]
Note that dequeue is only defined if the task space $t$ is at the head of the selected queue.

We briefly summarize each rule. \rulename{[Enqueue]} submits a task space to the runtime system on node 1.\footnote{A simplification; in actual implementations, tasks can begin on any node.}\rulename{[Distribute]} partitions a task space into smaller task spaces for mapping. \rulename{[Local]} selects the node where a task space's mapping decision will be made using {\tt SHARD}. \rulename{[Map]} assigns a task space to a specific processor within that node using the {\tt SLICE} function.

The semantics of task execution underscore the low-level nature of mapping in task-based systems. To achieve high performance, task spaces progress through multiple pipelined stages, each requiring user-defined mapping decisions. For brevity, we focus on {\tt SHARD} and {\tt SLICE}, though in practice, 19 distinct callback APIs are invoked across a task’s lifetime. This design fragments mapping logic across numerous low-level callbacks, making the APIs complex and opaque to most application developers.

\subsection{Translation}
\label{subsec:translationimpl}

The translation of \name{} code to the low-level runtime interface poses two key challenges. First, as discussed in \Cref{subsec:semantics}, Legion's mapping logic is spread across many callback functions triggered at different stages of a task's lifetime. While this structure mirrors the runtime's internal pipeline, it forces developers to work with a scattered and unintuitive interface (\Cref{fig:mappercompare}). Second, the low-level API lacks abstraction over the machine architecture, i.e., there is no logical view of the hardware, nor support for aligning mapping decisions with task spaces.

We implement \name{} by targeting Legion's programmatic mapping interface, comprising roughly 7,000 lines of C++ code—much of it dedicated to handling low-level runtime details. \name{} overrides a set of callback functions invoked at different stages of the task execution pipeline to decide task placement. The two core overrides, corresponding to the user-defined \texttt{SHARD} and \texttt{SLICE} functions, are illustrated in \Cref{fig:mappercompare} and explained in \Cref{subsec:semantics}.

We now give a concrete example using the \CodeIn{block1D\_x} distribution (\Cref{fig:codecommondist}). Let the processor space \CodeIn{m = Machine(GPU)} have shape $(n, g)$, representing $n$ nodes and $g$ GPUs per node. The mapper applies $m_1 = m.\mathtt{merge}(0,1).\mathtt{split}(0, ng)$, which reshapes the processor space to $(n g, 1)$. For an index point $(i_x, i_y)$ in a task space of size $(L_x, L_y)$, the \CodeIn{block1D\_x} function returns point $(\lfloor i_x \times \tfrac{n g}{L_x} \rfloor, 0)$ in $m_1$. The \name{} implementation then converts this point in $m_1$ back to a point in the original 2D processor space $m$ by applying the processor space transformations in reverse order (i.e., first inverting $\mathtt{split}(0,ng)$ and then inverting $\mathtt{merge}(0,1)$).
Thus, the $m_1$ index $(\lfloor i_x \times \tfrac{n g}{L_x} \rfloor,0)$ maps back to a point in $m$ as $(\text{node\_id}, \text{gpu\_id}) = (a \bmod n, \lfloor a / n \rfloor)$ where $a = \lfloor i_x \times \tfrac{n g}{L_x} \rfloor$. In the original space of $m$, $\text{node\_id}$ identifies the node and $\text{gpu\_id}$ specifies the GPU within that node. The node index is used by \CodeIn{SHARD}, applying either \rulename{[Distribute]} or \rulename{[Local]} depending on whether the target node is local, while the GPU index is used by \CodeIn{SLICE} under the \rulename{[Map]} rule.

\section{Results}
\label{sec:result}

Experiments were run on a cluster where each node has 40 IBM Power9 CPU cores and 4 NVIDIA V100 GPUs connected with NVLink 2.0 and an Infiniband EDR interconnect. To account for performance variability, each measurement was repeated five times, and the average is reported.

Our approach is evaluated on a suite of nine benchmarks chosen to demonstrate the generality and effectiveness of \name{} across representative distributed workloads. We use the published artifact repositories~\cite{yadav2022distal,bauer2012legion} and their corresponding problem sizes, where the mappers were developed by experts and tuned for the same machine, representing state-of-the-art implementations.

The first three benchmarks are drawn from scientific simulation workloads. Circuit~\citep{bauer2012legion} simulates electrical circuit behavior by modeling currents and voltages across interconnected nodes and wires. Stencil~\citep{stencilvan2014parallel} operates on a 2D grid, updating each point's value based on a stencil pattern derived from its neighbors. Pennant~\citep{ferenbaugh2015pennant} models unstructured mesh Lagrangian staggered-grid hydrodynamics, essential for simulating compressible flow.

The remaining six benchmarks are parallel matrix multiplication algorithms, an extensively studied problem in parallel computing that underpins many scientific applications. Matrix multiplication also serves as a classic stress test for distributed systems, as it requires carefully balancing computation and communication across multiple processors. They matrix multiplication algorithms we consider are \cannon~\citep{cannon1969cellular}, \summa~\citep{van1997summa}, \pumma~\citep{choi1994pumma}, \johnson~\citep{johnsonagarwal1995three}, \solomonik~\citep{solomonik2011communication}, and \cosma~\citep{cosmakwasniewski2019red}. They are divided into 2D and non-2D categories. The 2D algorithms, such as \cannon, \pumma, and \summa, enhance communication efficiency by partitioning matrices into 2D tiles and mapping them onto the processor space, utilizing techniques like pipelining. Conversely, non-2D algorithms like \johnson, \solomonik, and \cosma, employ 3D partitioning or optimize processor grids to minimize communication overhead and boost performance.

\subsection{Lines of Code Reduction}
\label{sec:result:locreduction}

We show the lines of code comparison between corresponding \name{} and C++ mappers in \Cref{tab:locreduction}. The application numbering in the \Cref{tab:locreduction} is consistent with the benchmark order introduced in the experimental setup of \Cref{sec:result}. For each mapper, we count the non-blank, non-comment lines of code for the \name{} mapper and the C++ mapper. Using \name{} to implement mappers leads to a reduction in lines of code from 406 to 29 on average, i.e., a \locreduction reduction across all benchmarks. In \Cref{fig:morefunctions} (see supplementary material), we show example functions used by mappers in distributed matrix multiplication. The DSL captures all mapping decisions of the C++ mappers succinctly.

We manually verify that both approaches produce \emph{identical} mapping decisions and \emph{matching performance} (i.e., identical throughput), showing that any overhead introduced by \name{} is negligible. This confirms the correctness and efficiency of our \name{} mappers relative to the original low-level C++ implementations in the published artifact repositories~\cite{yadav2022distal,bauer2012legion}.

We note that the application code is substantially larger than the mapping code and remains unchanged. The three scientific computing applications, Circuit, Stencil, and Pennant, are implemented in Regent~\cite{slaughter2015regent} with 776, 825, and 3.6K lines, respectively, and the six distributed matrix multiplication benchmarks each contain about 7.5K lines of Legion C++ code.

\begin{table*}[!tb]
\centering
\caption{Lines of Code (LoC) comparison between DSL and C++ mappers. The DSL achieves a \locreduction average reduction in LoC while maintaining the same performance as the C++ mappers.}
\postcaptionspace
\begin{tabular}{l|ccccccccc|c}
\toprule
Application & 1 & 2 & 3 & 4 & 5 & 6 & 7 & 8 & 9 & Avg. \\
\midrule 
LoC in C++ & 347 & 306 & 379 & 447 & 437 & 430 & 428 & 433 & 448 & 406 \\
LoC in \name{} & 16 & 14 & 16 & 38 & 38 & 38 & 33 & 38 & 32 & 29 \\ \midrule
LoC Reduction & $22\times$ & $22\times$ & $24\times$ & $12\times$ & $12\times$ & $11\times$ & $13\times$ & $11\times$ & $14\times$ & \locreduction \\
\bottomrule 
\end{tabular}
\label{tab:locreduction}
\end{table*}

\subsection{Performance Tuning}
\label{sec:result:performance}

We show that \name{} enables effective performance tuning by allowing users to create mappers that outperform existing expert-crafted C++ implementations. As shown in \Cref{tab:tuning}, where the application numbering follows the benchmark order introduced in the experimental setup of \Cref{sec:result}, \name{} achieves up to a 1.34$\times$ speedup over the baseline C++ mappers. These results demonstrate that \name{} helps developers identify mapping strategies that deliver higher throughput.

This performance gain comes from \name{}’s high-level declarative design, which is still expressive for high-performance mapping. Although our compiler translates \name{} mappers into standard C++ code, making the same strategies theoretically possible in C++, experts have not discovered them in practice. This is likely due to the low-level complexity of C++ mappers, where exploring new strategies often requires rewriting hundreds of lines of code. In contrast, similar changes in \name{} take only a few lines, making tuning much easier.

We further analyze the sources of performance gains. For the six matrix multiplication applications (indexed 4 to 9), improvements arise solely from optimized index mappings. For the three scientific applications, they result from different memory assignments enabled by \name{}’s full feature set (\Cref{sec:discussion}). Finally, case studies in \Cref{subsec:app:perf} (see supplementary material) illustrate how mapping choices impact performance.

\begin{table*}[!tb]
\centering
\caption{Performance improvements of \name{} mappers over expert-designed C++ mappers.}
\postcaptionspace
\begin{tabular}{l|ccccccccc}
\toprule
Application & 1 & 2 & 3 & 4 & 5 & 6 & 7 & 8 & 9 \\
\midrule 
\name{} Tuned Speedup & 1.34$\times$ & 1.02$\times$ & 1.04$\times$ & 1.09$\times$ & 1.09$\times$ & 1.09$\times$ & 1.09$\times$ & 1.07$\times$ & 1.31$\times$ \\
\bottomrule 
\end{tabular}
\label{tab:tuning}
\end{table*}

\begin{figure}[htbp]
\centering
    \begin{minipage}[t]{0.45\linewidth}
    \centering
        \includegraphics[width=0.8\linewidth]{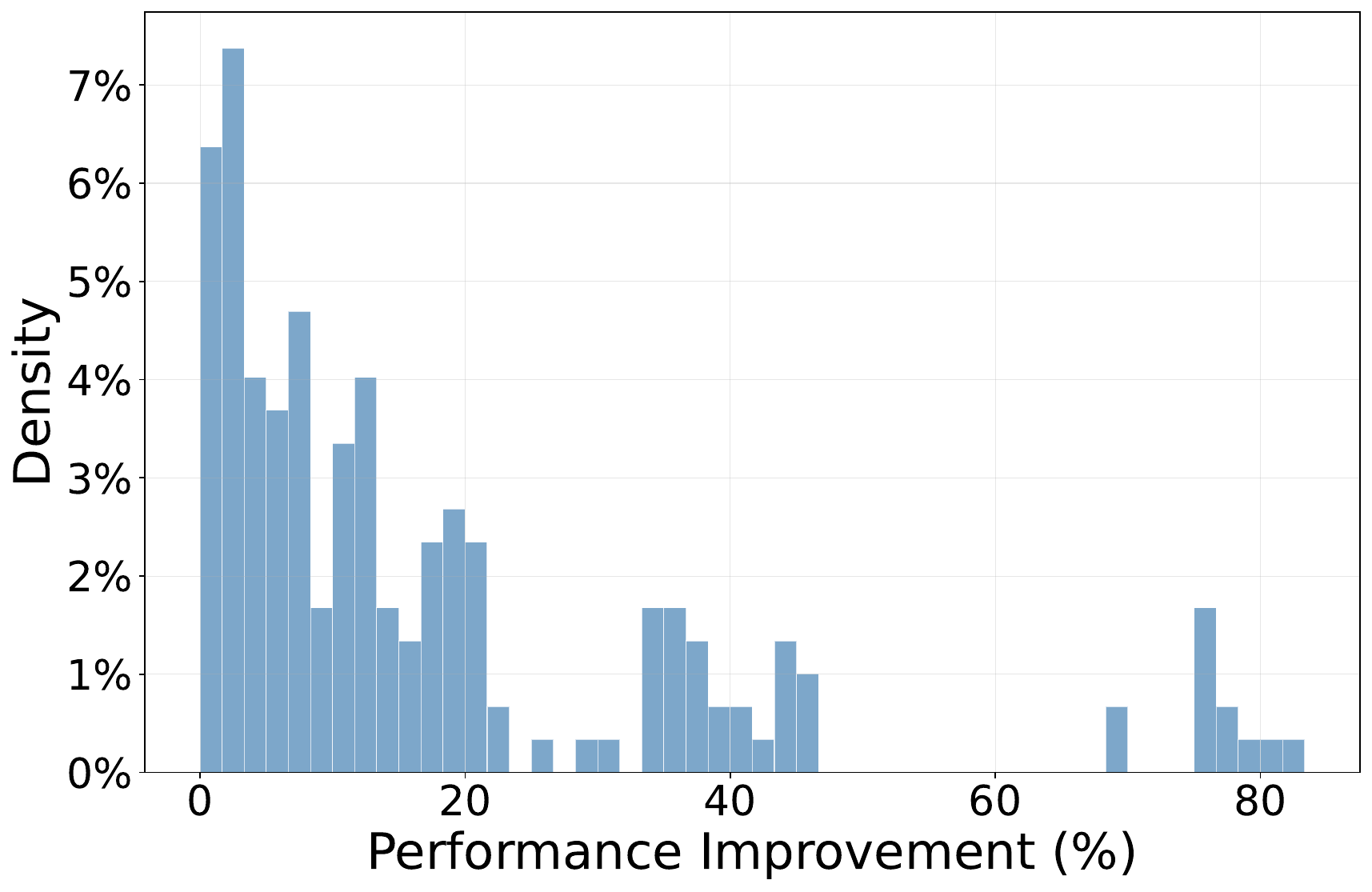}
        \vspace{-1em}
        \caption{The distribution of improvement percentage over different configurations.}
        \label{fig:distributionimprov}
    \end{minipage}%
    \hfill
    \begin{minipage}[t]{0.45\linewidth}
    \centering
        \includegraphics[width=0.8\linewidth]{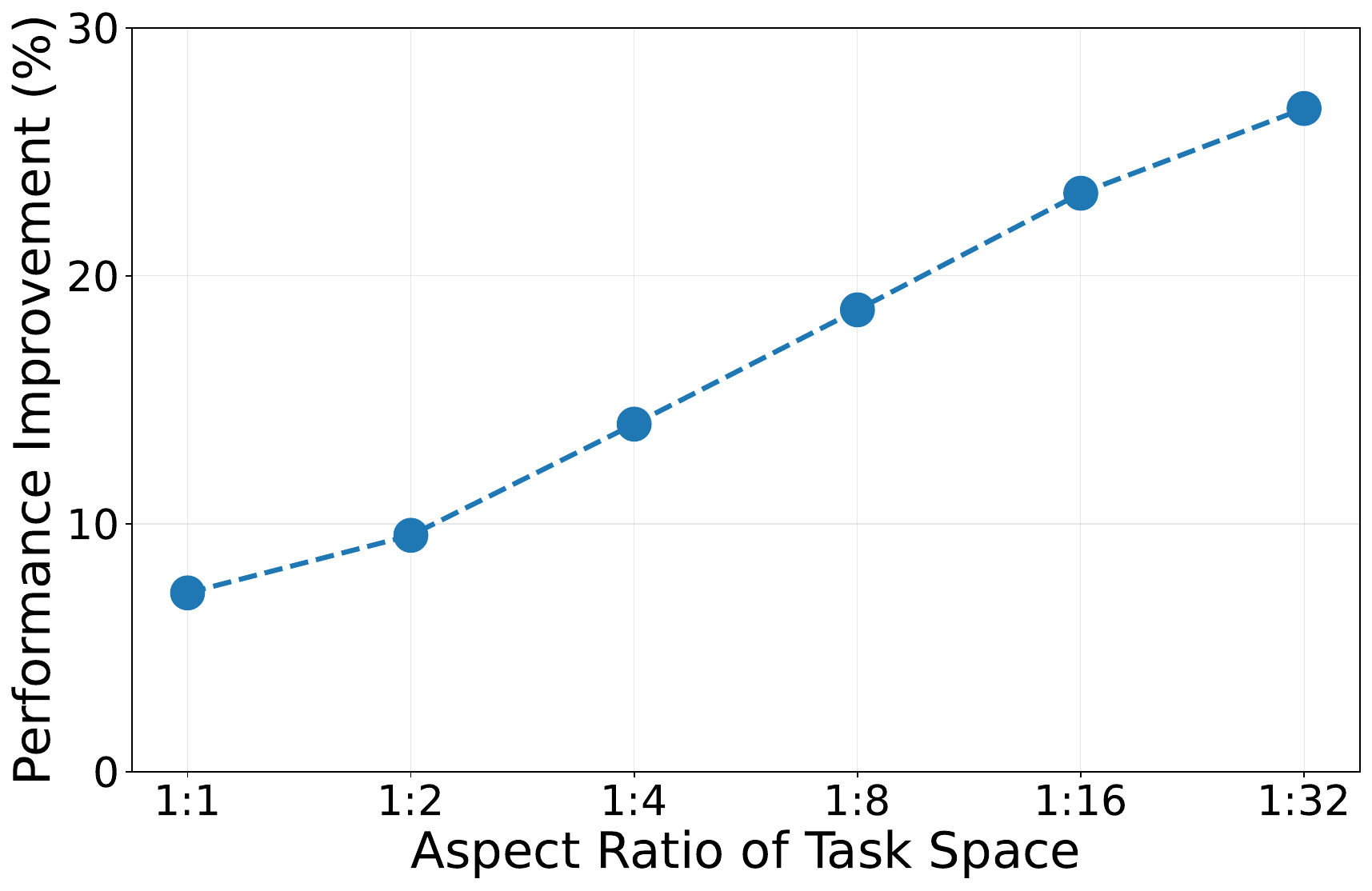}
        \vspace{-1em}
        \caption{Geometric mean of improvement percentage w.r.t. aspect ratios of task spaces.}
        \label{fig:aspectimprov}
    \end{minipage}
    \begin{minipage}[t]{0.45\linewidth}
    \centering
        \includegraphics[width=0.8\linewidth]{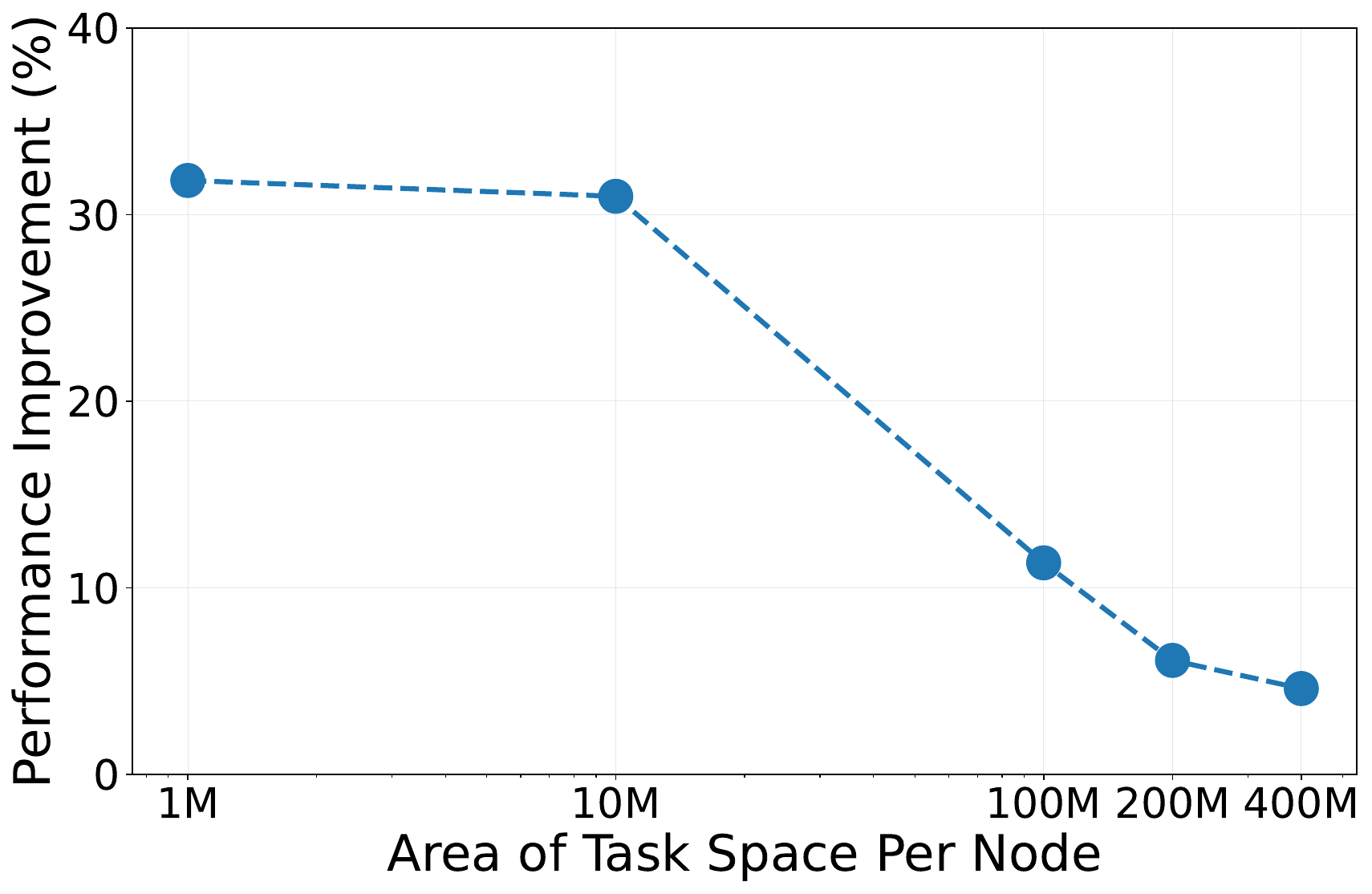}
        \vspace{-1em}
        \caption{Geometric mean of improvement percentage w.r.t. area of task space per node.}
        \label{fig:areaimprov}
    \end{minipage}
    \hfill
    \begin{minipage}[t]{0.45\linewidth}
    \centering
        \includegraphics[width=0.8\linewidth]{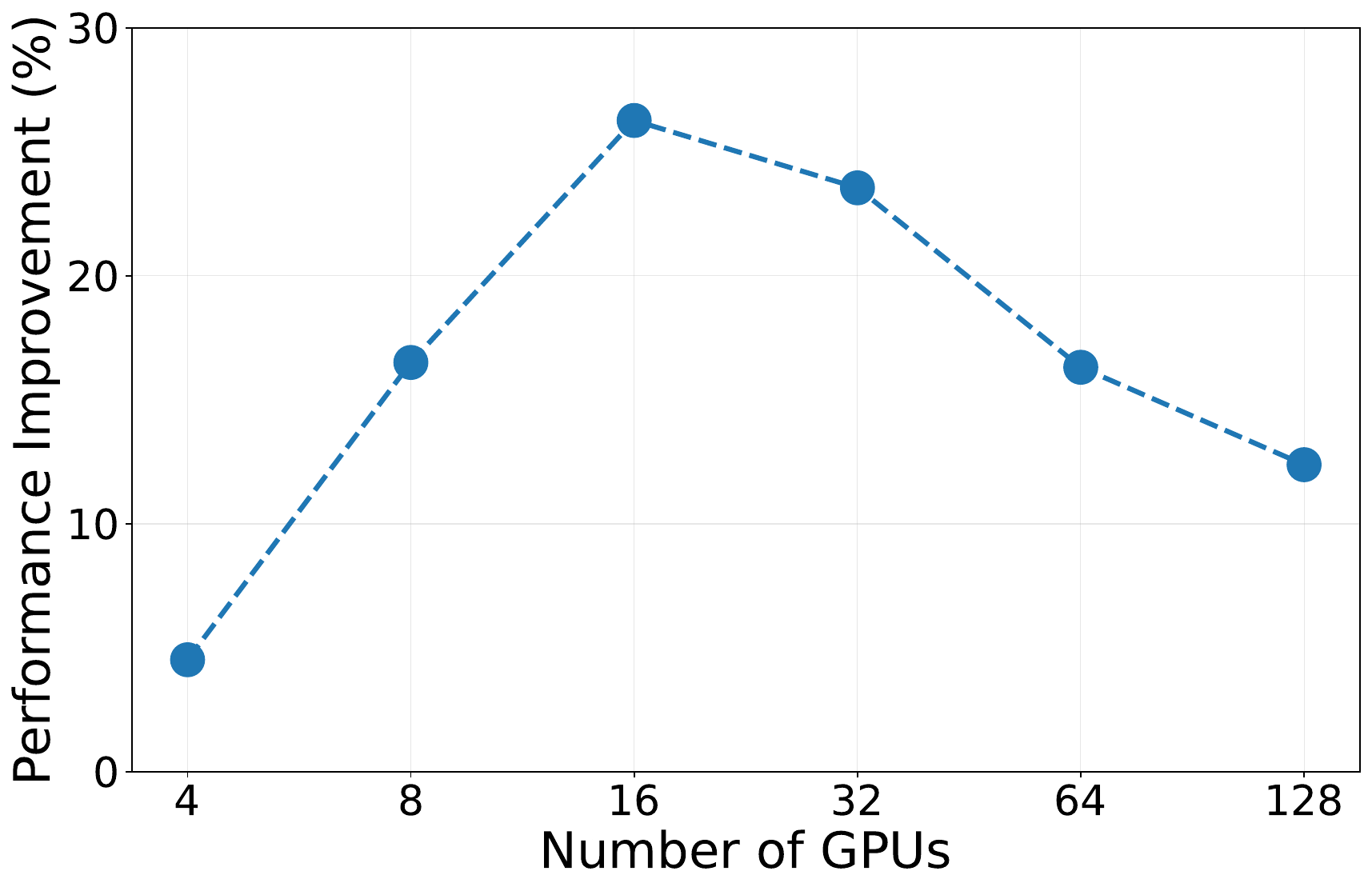}
        \vspace{-1em}
        \caption{Geometric mean of improvement percentage w.r.t. machine sizes.}
        \label{fig:machineimprov}
    \end{minipage}
\precaptionspace
\end{figure}

\subsection{Performance Study of the Decompose Primitive}
\label{sec:result:primitive}

To evaluate the impact of the \CodeIn{decompose} primitive, we compare it with the default machine space reshaping heuristic from \Cref{alg:chapel} (see supplementary material). We measure end-to-end performance on stencil applications, which are common in image processing and scientific computing, where each point updates based on its neighbors. As these applications are sensitive to data partitioning and mapping, the results provide strong evidence of \CodeIn{decompose} improving communication efficiency in practical workloads.

To ensure a fair comparison, we systematically vary the 2D stencil parameters. We test six aspect ratios ($1:1$, $1:2$, $1:4$, $1:8$, $1:16$, $1:32$), representative of shapes typical in scientific simulations (e.g., liquid films~\cite{nave2010direct}, shock tubes~\cite{wong2019high}). We also vary the task space size per node across five scales ($10^6$, $10^7$, $10^8$, $2{\times}10^8$, $4{\times}10^8$) for different communication-to-computation ratios, and scale the number of GPUs from 4 to 128 in powers of two. In total, we evaluate 180 configurations.

We report percentage improvements in \Cref{fig:distributionimprov}, ranging from 0\% to \improvemax{}, with a geometric mean of \improveavg{}. \Cref{fig:aspectimprov} plots the geometric mean improvement against the task space aspect ratio. As the ratio increases from $1:1$ to $1:32$, the improvement rises from 7\% to 27\%, showing the growing benefit of using the \CodeIn{decompose} primitive. This trend is expected since \Cref{alg:chapel} evenly splits processors, which works best when the task space is nearly square.

\Cref{fig:areaimprov} shows how improvement varies with task space size per node. As the area grows from $10^6$ to $4 \times 10^8$, the improvement decreases from 32\% to 5\%, since larger spaces lower the communication-to-computation ratio. \Cref{fig:machineimprov} plots improvement against machine size. The benefit peaks at 26\% on 4 nodes: increasing from 1 to 4 nodes amplifies inter-node communication, while beyond 4 nodes, inter-rack latency becomes the bottleneck, reducing the gain.

\section{Discussion}
\label{sec:discussion}

\begin{figure*}[t]
\footnotesize
\setlength{\jot}{2pt}

\begin{minipage}[t]{0.42\linewidth}
\begin{align*}
\text{Program}\; &::=\; \text{Statement}^+ \\[0.2em]
\text{Statement}\; &::=\; \text{TaskMap} \mid \text{DataMap} \mid \text{DataLayout} \mid {}\\
&\qquad \text{FuncDef} \mid \kw{IndexTaskMap}\;\mathit{t}\;\mathit{v} \mid \cdots \\[0.2em]
\text{TaskMap}\; &::=\; \kw{Task}\;\mathit{t}\;\text{Proc}^+ \\[0.2em]
\text{DataMap}\; &::=\; \kw{Region}\;\mathit{t}\;\mathit{r}\;\text{Proc}\;\text{Memory}^+ \\[0.2em]
\text{Proc}\; &::=\; \kw{CPU} \mid \kw{GPU} \mid \kw{OMP} \\[0.2em]
\text{Memory}\; &::=\; \kw{SYSMEM} \mid \kw{FBMEM} \mid \kw{ZCMEM} \\[0.2em]
\text{DataLayout}\; &::=\; \kw{Layout}\;\mathit{t}\;\mathit{r}\;\text{Proc}\;\text{Constraint}^+ \\[0.2em]
\end{align*}
\end{minipage}\hfill
\begin{minipage}[t]{0.42\linewidth}
\begin{align*}
\text{Constraint}\; &::=\; \kw{SOA} \mid \kw{AOS} \mid \kw{C\_order} \mid \kw{F\_order} \mid \kw{Align}{\ ==\ } \mathit{i} \\[0.2em]
\text{FuncDef}\; &::=\; \kw{def}\;\mathit{v}(\mathit{v}^+):\;\text{FuncStmt}^+ \\[0.2em]
\text{FuncStmt}\; &::=\; \mathit{v} \;{=}\; \text{Expr} \mid \kw{return}\;\text{Expr} \\[0.3em]
\text{Expr}\; &::=\; \mathit{v} \mid \mathit{v}(\text{Expr}^+) \mid \kw{Machine}(\text{Proc}) \mid {}\\
&\qquad \text{Expr}\mathbin{.}\text{Expr} \mid \text{Expr}\;\kw{Op}\;\text{Expr} \mid (\text{Expr}) \mid {}\\
&\qquad \text{Expr [Expr]} \mid {*}\ \text{Expr} \mid \text{Expr}\; ?\; \text{Expr}\; :\; \text{Expr} \mid {}\\
&\qquad \text{Primitive} \\[0.3em]
\text{Primitive}\; &::=\; \kw{split} \mid \kw{merge} \mid \kw{swap} \mid \kw{slice} \mid \kw{decompose}
\end{align*}
\end{minipage}

\vspace{-1em}
\noindent\small
$\mathit{t}\in\text{\textbf{Task Names}}\quad
\mathit{r}\in\text{\textbf{Region Names}}\quad
\mathit{v}\in\text{\textbf{Variables}}\quad
\mathit{i}\in\text{\textbf{Integers}}$
\caption{Syntax of the \name{} language.}
\label{fig:dslgrammar}
\end{figure*}

While \Cref{sec:method} and \Cref{sec:decompose} focus on index mapping, the full optimization space includes several additional dimensions that contribute to the performance gains observed in the three scientific applications evaluated in \Cref{sec:result:performance}. \Cref{fig:dslgrammar} presents a simplified grammar capturing the core constructs of \name{}. A primary axis of control is \emph{processor selection}, expressed via the \CodeIn{TaskMap} directive. This determines whether a given task is assigned to GPUs, CPUs, or the OpenMP runtime. The decision is made per task and depends on factors such as task granularity, GPU memory capacity, and kernel launch overhead. For example, small tasks may favor CPUs to avoid the cost of GPU launches, and large-memory tasks may also favor CPUs if they exceed GPU memory limits.

Another critical dimension is \emph{memory placement}, specified through the \CodeIn{DataMap} construct. This governs where task arguments are stored: in GPU FrameBuffer memory for high-speed access, in ZeroCopy regions to enable CPU-GPU sharing, or in host memory when capacity is a concern. Each location involves trade-offs between access latency, available memory, and transfer costs. This mapping is defined per task and per argument.

The optimization space also includes \emph{memory layout}, configured with the \CodeIn{DataLayout} statement. This involves selecting between layouts such as Struct of Arrays (SoA) versus Array of Structures (AoS), specifying memory ordering (e.g., Fortran vs. C order), and applying alignment constraints (e.g., 128-byte alignment). These layout choices directly impact cache behavior and performance. This decision is made per task, per data region, and per target processor. Additional features, not shown in the figure, include scheduling policies (e.g., task prioritization), garbage collection strategies, and load-balancing directives.

While \name{} captures a broad range of optimization dimensions, it is intentionally constrained to focus on performance-critical mapping decisions that, in our experience, are the most common in practice. One current limitation is that \name{} is stateless, making it less suitable for highly dynamic workloads or computations that would benefit from retaining historical information across mapping decisions. Although it could be extended with a persistent state, we have not yet found this necessary. Beyond this, we have not encountered cases where \name{} is less expressive than the C++ interface. Notably, every mapper that achieved higher performance than the expert-written C++ implementations was expressible in \name{}, demonstrating that \name{} is effective despite being more restrictive than general-purpose C++, at least for the applications we studied.

\section{Related Work}
\label{sec:related}

\subsection{Task-based Programming Systems}

\begin{table*}
\caption{Mapping features exposed to application control by different systems and covered by Mapple.}
\postcaptionspace
\centering    
\scalebox{0.8}{
\begin{tabular}{l|ccccccc}\hline
Systems &Task Placement & Data Placement & Data Layout & Scheduling &Load Balancing \\\hline
\legion~\cite{bauer2012legion} &\checkmark &\checkmark & \checkmark &\checkmark &\checkmark \\
\starpu~\cite{augonnet2010starpu} &\checkmark & \checkmark  &\checkmark &\checkmark &\checkmark \\
\chapel~\cite{chamberlain2007parallelchapel} &\checkmark & & \checkmark & &\checkmark \\
\charmpp~\cite{kale1993charm++} &\checkmark &\checkmark & &\checkmark &\checkmark \\
\parsec~\cite{danalis2015parsec} &\checkmark & & \checkmark &\checkmark &\checkmark \\
\xten~\cite{charles2005x10} &\checkmark &\checkmark & &\checkmark &\checkmark \\
\hpx~\cite{heller2017hpx} &\checkmark &\checkmark & &\checkmark &\checkmark \\
\openmp~\cite{chandra2001parallelopenmp} &\checkmark &  & & & \checkmark \\
\ompss~\cite{duran2011ompss} &\checkmark & & &\checkmark &\checkmark \\
\hline
\textbf{\name{}} &\checkmark &\checkmark &\checkmark &\checkmark &\checkmark \\
\hline
\end{tabular}
}
\label{tab:survey}
\end{table*}

Table~\ref{tab:survey} summarizes mapping features in HPC systems, all of which are supported by \name{}. Systems like Legion~\cite{bauer2012legion} and StarPU~\cite{augonnet2009starpu} offer low-level C/C++ mapping interfaces via asynchronous callbacks, requiring deep familiarity with internal runtime abstractions. Other HPC systems offer less direct mapping control. ZPL~\cite{deitz2004abstractionszpl}, Chapel~\cite{chamberlain2007parallelchapel}, and \xten~\cite{charles2005x10} expose predefined data distributions (e.g., blocked, cyclic). HPX~\cite{heller2017hpx} and \parsec~\cite{danalis2015parsec} rely on schedulers with tunable policies for heterogeneous execution. OpenMP~\cite{chandra2001parallelopenmp}, \ompss~\cite{duran2011ompss}, \charmpp{}~\cite{kale1993charm++}, and Chapel's low-level APIs allow programmers to specify task placement directly in the source code, ranging from processor annotations (OpenMP, \ompss{}) to custom C++ mapping classes (\charmpp{}). Task-based systems in data analytics and AI such as \mapreduce{}~\cite{dean2008mapreduce}, \spark{}~\cite{zaharia2010spark}, \dask{}~\cite{rocklin2015dask}, \ray{}~\cite{moritz2018ray}, and \pathways{}~\cite{barham2022pathways} rely on runtime mapping heuristics. This eases development but limits performance tuning when heuristics fall short. \name{} offers better ease-of-use than low-level mapping interfaces while enabling finer-grained control than systems that rely entirely on heuristics built into the runtime system.

\subsection{Loop Transformations}

A large body of work on compiler algorithms uses loop transformations to maximize parallelism and minimize communication, including unimodular transformations~\cite{banerjee1990unimodular,banerjee2007loop,wolf1992improving,wolf1991loop,wolfe1982optimizing} (i.e., combinations of loop interchange, reversal, and skewing), loop block/tiling, fusion, fission, reindexing, and scaling~\cite{lim1999affine,lim2001blocking,wolf1991data,xue2000loop}, and affine transformations used by the polyhedral model~\cite{bondhugula2013compiling,bondhugula2008practical,acharya2018polyhedral}. These works share a similar motivation to \name{}, though the goals and specific transformations are different as explained in \Cref{sec:intro}.

Another line of work uses scheduling languages~\cite{chen2008framework,ragan2012decoupling,zhang2018graphit,baghdadi2019tiramisu,kjolstad2017tensor,senanayake2020sparse,chen2018tvm,yadav2022spdistal} to separate the algorithm from the schedule, based on traditional loop optimizations such as split, collapse, and loop reordering. This separation enables programmers to just change the schedule when moving to different hardware. A growing body of research focuses on auto-scheduling~\cite{adams2019learning,chen2018learning,mullapudi2016automatically,zheng2020ansor,zheng2022amos,tollenaere2023autotuning} to reduce the manual effort required in writing schedules. Task-based systems share a similar philosophy of separating the algorithm from its schedule or mapping. However, prior scheduling languages such as DISTAL~\cite{yadav2022spdistal} and Distributed Halide~\cite{denniston2016distributed} are specialized compiler frameworks for tensor algebra that lower high-level tensor expressions to distributed execution. In contrast, \name{} addresses the general mapping problem in task-based runtime systems (see \Cref{sec:discussion}), supporting a broader class of task-based programs. Its design centers on the \emph{decompose} primitive (see \Cref{sec:decompose}), which automatically resolves dimensionality mismatches between task and processor spaces. In our evaluation, DISTAL’s C++ mappers serve as baselines; \name{} reduces their code size by over 11× and discovers mappers that achieve higher performance.

\section{Conclusion}
\label{sec:conclusion}

We present \name{}, a high-level interface that simplifies the development of mappers for task-based parallel systems. \name{} overcomes key limitations of existing mapping interfaces by abstracting low-level runtime details while exposing transformation primitives to resolve dimensionality mismatches between task and processor spaces. At the core of \name{} is the \emph{decompose} primitive, which helps minimize communication volume. Across nine distributed applications, \name{} reduces mapper code size by \locreduction{} and achieves up to \mapperspeedup{} speedup over expert-written C++ mappers. Moreover, the \emph{decompose} primitive outperforms existing heuristics by up to \primitivespeedup. These results demonstrate the practical effectiveness of \name{} in enabling the development of concise, high-performance mappers for distributed applications.


\bibliographystyle{ACM-Reference-Format}
\bibliography{custom}

@String{Computing = "Computing" }

@String{Computer = "{IEEE} Computer" }

@String{Springer = "Springer-Verlag" }

@inproceedings{chamberlain2011authoring,
  title={Authoring user-defined domain maps in Chapel},
  author={Chamberlain, Bradford L and Choi, Sung-Eun and Deitz, Steven J and Iten, David and Litvinov, Vassily},
  booktitle={Cray Users Group Conference (CUG)},
  year={2011}
}

@book{sanders2010cuda,
  title={CUDA by example: an introduction to general-purpose GPU programming},
  author={Sanders, Jason and Kandrot, Edward},
  year={2010},
  publisher={Addison-Wesley Professional}
}

@book{chandra2001parallel,
  title={Parallel programming in OpenMP},
  author={Chandra, Rohit},
  year={2001},
  publisher={Morgan kaufmann}
}

@misc{ampi,
    title = {{Adaptive MPI}},
    key = {{Adaptive MPI}},
    year = {2025},
    note = {\url{https://charm.readthedocs.io/en/latest/ampi/04-extensions.html##user-defined-initial-mapping}}
}

@misc{keyword,
    title = {{Keyword Distribution}},
    key = {{Keyword Distribution}},
    year = {2025},
    note = {\url{https://chapel-lang.org/docs/1.28/modules/dists/BlockDist.html}}
}

@article{dagum1998openmp,
  title={OpenMP: an industry standard API for shared-memory programming},
  author={Dagum, Leonardo and Menon, Ramesh},
  journal={IEEE computational science and engineering},
  volume={5},
  number={1},
  pages={46--55},
  year={1998},
  publisher={IEEE}
}

@article{yang2011hybrid,
  title={Hybrid CUDA, OpenMP, and MPI parallel programming on multicore GPU clusters},
  author={Yang, Chao-Tung and Huang, Chih-Lin and Lin, Cheng-Fang},
  journal={Computer Physics Communications},
  volume={182},
  number={1},
  pages={266--269},
  year={2011},
  publisher={Elsevier}
}

@book{gropp1999using,
  title={Using MPI: portable parallel programming with the message-passing interface},
  author={Gropp, William and Lusk, Ewing and Skjellum, Anthony},
  volume={1},
  year={1999},
  publisher={MIT press}
}

@inproceedings{slaughter2015regent,
  title={Regent: A high-productivity programming language for HPC with logical regions},
  author={Slaughter, Elliott and Lee, Wonchan and Treichler, Sean and Bauer, Michael and Aiken, Alex},
  booktitle={Proceedings of the International Conference for High Performance Computing, Networking, Storage and Analysis},
  pages={1--12},
  year={2015}
}

@inproceedings{iverson1962programming,
  title={A programming language},
  author={Iverson, Kenneth E},
  booktitle={Proceedings of the May 1-3, 1962, spring joint computer conference},
  pages={345--351},
  year={1962}
}

@article{denniston2016distributed,
  title={Distributed halide},
  author={Denniston, Tyler and Kamil, Shoaib and Amarasinghe, Saman},
  journal={ACM SIGPLAN Notices},
  volume={51},
  number={8},
  pages={1--12},
  year={2016},
  publisher={ACM New York, NY, USA}
}

@inproceedings{bauer2012legion,
  title={Legion: Expressing locality and independence with logical regions},
  author={Bauer, Michael and Treichler, Sean and Slaughter, Elliott and Aiken, Alex},
  booktitle={SC'12: Proceedings of the International Conference on High Performance Computing, Networking, Storage and Analysis},
  pages={1--11},
  year={2012},
  organization={IEEE}
}

@phdthesis{augonnet2010starpu,
  title={StarPU: a runtime system for scheduling tasks over accelerator-based multicore machines},
  author={Augonnet, C{\'e}dric and Thibault, Samuel and Namyst, Raymond},
  year={2010},
  school={INRIA}
}

@inproceedings{augonnet2009starpu,
  title={StarPU: a unified platform for task scheduling on heterogeneous multicore architectures},
  author={Augonnet, C{\'e}dric and Thibault, Samuel and Namyst, Raymond and Wacrenier, Pierre-Andr{\'e}},
  booktitle={European Conference on Parallel Processing},
  pages={863--874},
  year={2009},
  organization={Springer}
}

@inproceedings{danalis2015parsec,
  title={Parsec in practice: Optimizing a legacy chemistry application through distributed task-based execution},
  author={Danalis, Anthony and Jagode, Heike and Bosilca, George and Dongarra, Jack},
  booktitle={2015 IEEE International Conference on Cluster Computing},
  pages={304--313},
  year={2015},
  organization={IEEE}
}

@inproceedings{chamberlain2010user,
  title={User-defined distributions and layouts in Chapel: Philosophy and framework},
  author={Chamberlain, Bradford L and Deitz, Steven J and Iten, David and Choi, Sung-Eun},
  booktitle={Proceedings of the 2nd USENIX conference on Hot topics in parallelism},
  pages={12--12},
  year={2010}
}

@article{barham2022pathways,
  title={Pathways: Asynchronous distributed dataflow for ML},
  author={Barham, Paul and Chowdhery, Aakanksha and Dean, Jeff and Ghemawat, Sanjay and Hand, Steven and Hurt, Daniel and Isard, Michael and Lim, Hyeontaek and Pang, Ruoming and Roy, Sudip and others},
  journal={Proceedings of Machine Learning and Systems},
  volume={4},
  pages={430--449},
  year={2022}
}

@inproceedings{moritz2018ray,
  title={Ray: A distributed framework for emerging $\{$AI$\}$ applications},
  author={Moritz, Philipp and Nishihara, Robert and Wang, Stephanie and Tumanov, Alexey and Liaw, Richard and Liang, Eric and Elibol, Melih and Yang, Zongheng and Paul, William and Jordan, Michael I and others},
  booktitle={13th USENIX Symposium on Operating Systems Design and Implementation (OSDI 18)},
  pages={561--577},
  year={2018}
}

@inproceedings{rocklin2015dask,
  title={Dask: Parallel computation with blocked algorithms and task scheduling},
  author={Rocklin, Matthew},
  booktitle={Proceedings of the 14th python in science conference},
  volume={130},
  pages={136},
  year={2015},
  organization={Citeseer}
}

@article{chamberlain2007parallelchapel,
  title={Parallel programmability and the chapel language},
  author={Chamberlain, Bradford L and Callahan, David and Zima, Hans P},
  journal={The International Journal of High Performance Computing Applications},
  volume={21},
  number={3},
  pages={291--312},
  year={2007},
  publisher={Sage Publications Sage UK: London, England}
}

@inproceedings{fatahalian2006sequoia,
  title={Sequoia: Programming the memory hierarchy},
  author={Fatahalian, Kayvon and Horn, Daniel Reiter and Knight, Timothy J and Leem, Larkhoon and Houston, Mike and Park, Ji Young and Erez, Mattan and Ren, Manman and Aiken, Alex and Dally, William J and others},
  booktitle={Proceedings of the 2006 ACM/IEEE Conference on Supercomputing},
  pages={83--es},
  year={2006}
}

@book{chandra2001parallelopenmp,
  title={Parallel programming in OpenMP},
  author={Chandra, Rohit and Dagum, Leo and Kohr, David and Menon, Ramesh and Maydan, Dror and McDonald, Jeff},
  year={2001},
  publisher={Morgan kaufmann}
}

@article{duran2011ompss,
  title={Ompss: a proposal for programming heterogeneous multi-core architectures},
  author={Duran, Alejandro and Ayguad{\'e}, Eduard and Badia, Rosa M and Labarta, Jes{\'u}s and Martinell, Luis and Martorell, Xavier and Planas, Judit},
  journal={Parallel processing letters},
  volume={21},
  number={02},
  pages={173--193},
  year={2011},
  publisher={World Scientific}
}

@inproceedings{kale1993charm++,
  title={Charm++ a portable concurrent object oriented system based on c++},
  author={Kale, Laxmikant V and Krishnan, Sanjeev},
  booktitle={Proceedings of the eighth annual conference on Object-oriented programming systems, languages, and applications},
  pages={91--108},
  year={1993}
}

@inproceedings{kaiser2014hpx,
  title={Hpx: A task based programming model in a global address space},
  author={Kaiser, Hartmut and Heller, Thomas and Adelstein-Lelbach, Bryce and Serio, Adrian and Fey, Dietmar},
  booktitle={Proceedings of the 8th International Conference on Partitioned Global Address Space Programming Models},
  pages={1--11},
  year={2014}
}

@article{heller2017hpx,
  title={Hpx--an open source c++ standard library for parallelism and concurrency},
  author={Heller, Thomas and Diehl, Patrick and Byerly, Zachary and Biddiscombe, John and Kaiser, Hartmut},
  journal={Proceedings of OpenSuCo},
  volume={5},
  year={2017}
}

@book{cannon1969cellular,
  title={A cellular computer to implement the Kalman filter algorithm},
  author={Cannon, Lynn Elliot},
  year={1969},
  publisher={Montana State University}
}

@article{choi1994pumma,
  title={PUMMA: Parallel universal matrix multiplication algorithms on distributed memory concurrent computers},
  author={Choi, Jaeyoung and Walker, David W and Dongarra, Jack J},
  journal={Concurrency: Practice and Experience},
  volume={6},
  number={7},
  pages={543--570},
  year={1994},
  publisher={Wiley Online Library}
}

@article{van1997summa,
  title={SUMMA: Scalable universal matrix multiplication algorithm},
  author={Van De Geijn, Robert A and Watts, Jerrell},
  journal={Concurrency: Practice and Experience},
  volume={9},
  number={4},
  pages={255--274},
  year={1997},
  publisher={Wiley Online Library}
}

@article{johnsonagarwal1995three,
  title={A three-dimensional approach to parallel matrix multiplication},
  author={Agarwal, Ramesh C and Balle, Susanne M and Gustavson, Fred G and Joshi, Mahesh and Palkar, Prasad},
  journal={IBM Journal of Research and Development},
  volume={39},
  number={5},
  pages={575--582},
  year={1995},
  publisher={IBM}
}

@inproceedings{solomonik2011communication,
  title={Communication-optimal parallel 2.5 D matrix multiplication and LU factorization algorithms},
  author={Solomonik, Edgar and Demmel, James},
  booktitle={Euro-Par 2011 Parallel Processing: 17th International Conference, Euro-Par 2011, Bordeaux, France, August 29-September 2, 2011, Proceedings, Part II 17},
  pages={90--109},
  year={2011},
  organization={Springer}
}

@inproceedings{cosmakwasniewski2019red,
  title={Red-blue pebbling revisited: near optimal parallel matrix-matrix multiplication},
  author={Kwasniewski, Grzegorz and Kabi{\'c}, Marko and Besta, Maciej and VandeVondele, Joost and Solc{\`a}, Raffaele and Hoefler, Torsten},
  booktitle={Proceedings of the International Conference for High Performance Computing, Networking, Storage and Analysis},
  pages={1--22},
  year={2019}
}

@article{ferenbaugh2015pennant,
  title={PENNANT: an unstructured mesh mini-app for advanced architecture research},
  author={Ferenbaugh, Charles R},
  journal={Concurrency and Computation: Practice and Experience},
  volume={27},
  number={17},
  pages={4555--4572},
  year={2015},
  publisher={Wiley Online Library}
}

@inproceedings{stencilvan2014parallel,
  title={The parallel research kernels},
  author={Van der Wijngaart, Rob F and Mattson, Timothy G},
  booktitle={2014 IEEE High Performance Extreme Computing Conference (HPEC)},
  pages={1--6},
  year={2014},
  organization={IEEE}
}

@inproceedings{yadav2022distal,
  title={DISTAL: the distributed tensor algebra compiler},
  author={Yadav, Rohan and Aiken, Alex and Kjolstad, Fredrik},
  booktitle={Proceedings of the 43rd ACM SIGPLAN International Conference on Programming Language Design and Implementation},
  pages={286--300},
  year={2022}
}

@article{nave2010direct,
  title={Direct numerical simulation of liquid films with large interfacial deformation},
  author={Nave, J-C and Liu, XD and Banerjee, Sanjoy},
  journal={Studies in Applied Mathematics},
  volume={125},
  number={2},
  pages={153--177},
  year={2010},
  publisher={Wiley Online Library}
}

@article{wong2019high,
  title={High-resolution Navier-Stokes simulations of Richtmyer-Meshkov instability with reshock},
  author={Wong, Man Long and Livescu, Daniel and Lele, Sanjiva K},
  journal={Physical Review Fluids},
  volume={4},
  number={10},
  pages={104609},
  year={2019},
  publisher={APS}
}

@book{banerjee1990unimodular,
  title={Unimodular transformations of double loops},
  author={Banerjee, Utpal and others},
  year={1990},
  publisher={University of Illinois at Urbana-Champaign, Center for Supercomputing~…}
}

@book{banerjee2007loop,
  title={Loop transformations for restructuring compilers: the foundations},
  author={Banerjee, Utpal},
  year={2007},
  publisher={Springer Science \& Business Media}
}

@book{wolf1992improving,
  title={Improving locality and parallelism in nested loops},
  author={Wolf, Michael Edward},
  year={1992},
  publisher={stanford university}
}

@article{wolf1991loop,
  title={A loop transformation theory and an algorithm to maximize parallelism},
  author={Wolf, Michael E and Lam, Monica S},
  journal={IEEE Transactions on Parallel \& Distributed Systems},
  volume={2},
  number={04},
  pages={452--471},
  year={1991},
  publisher={IEEE Computer Society}
}

@book{wolfe1982optimizing,
  title={Optimizing supercompilers for supercomputers},
  author={Wolfe, Michael Joseph},
  year={1982},
  publisher={University of Illinois at Urbana-Champaign}
}

@inproceedings{lim1999affine,
  title={An affine partitioning algorithm to maximize parallelism and minimize communication},
  author={Lim, Amy W and Cheong, Gerald I and Lam, Monica S},
  booktitle={Proceedings of the 13th international conference on Supercomputing},
  pages={228--237},
  year={1999}
}

@inproceedings{lim2001blocking,
  title={Blocking and array contraction across arbitrarily nested loops using affine partitioning},
  author={Lim, Amy W and Liao, Shih-Wei and Lam, Monica S},
  booktitle={Proceedings of the eighth ACM SIGPLAN symposium on Principles and practices of parallel programming},
  pages={103--112},
  year={2001}
}

@inproceedings{bondhugula2013compiling,
  title={Compiling affine loop nests for distributed-memory parallel architectures},
  author={Bondhugula, Uday},
  booktitle={Proceedings of the International Conference on High Performance Computing, Networking, Storage and Analysis},
  pages={1--12},
  year={2013}
}

@inproceedings{bondhugula2008practical,
  title={A practical automatic polyhedral parallelizer and locality optimizer},
  author={Bondhugula, Uday and Hartono, Albert and Ramanujam, Jagannathan and Sadayappan, Ponnuswamy},
  booktitle={Proceedings of the 29th ACM SIGPLAN Conference on Programming Language Design and Implementation},
  pages={101--113},
  year={2008}
}

@inproceedings{wolf1991data,
  title={A data locality optimizing algorithm},
  author={Wolf, Michael E and Lam, Monica S},
  booktitle={Proceedings of the ACM SIGPLAN 1991 conference on Programming language design and implementation},
  pages={30--44},
  year={1991}
}

@book{xue2000loop,
  title={Loop tiling for parallelism},
  author={Xue, Jingling},
  volume={575},
  year={2000},
  publisher={Springer Science \& Business Media}
}

@article{chen2008framework,
  title={A framework for composing high-level loop transformations},
  author={Chen, Chun and Chame, Jacqueline and Hall, Mary},
  journal={Technical Report 08--897, USC Computer Science Technical Report},
  year={2008}
}

@article{ragan2012decoupling,
  title={Decoupling algorithms from schedules for easy optimization of image processing pipelines},
  author={Ragan-Kelley, Jonathan and Adams, Andrew and Paris, Sylvain and Levoy, Marc and Amarasinghe, Saman and Durand, Fr{\'e}do},
  journal={ACM Transactions on Graphics (TOG)},
  volume={31},
  number={4},
  pages={1--12},
  year={2012},
  publisher={ACM New York, NY, USA}
}

@article{zhang2018graphit,
  title={Graphit: A high-performance graph dsl},
  author={Zhang, Yunming and Yang, Mengjiao and Baghdadi, Riyadh and Kamil, Shoaib and Shun, Julian and Amarasinghe, Saman},
  journal={Proceedings of the ACM on Programming Languages},
  volume={2},
  number={OOPSLA},
  pages={1--30},
  year={2018},
  publisher={ACM New York, NY, USA}
}

@inproceedings{baghdadi2019tiramisu,
  title={Tiramisu: A polyhedral compiler for expressing fast and portable code},
  author={Baghdadi, Riyadh and Ray, Jessica and Romdhane, Malek Ben and Del Sozzo, Emanuele and Akkas, Abdurrahman and Zhang, Yunming and Suriana, Patricia and Kamil, Shoaib and Amarasinghe, Saman},
  booktitle={2019 IEEE/ACM International Symposium on Code Generation and Optimization (CGO)},
  pages={193--205},
  year={2019},
  organization={IEEE}
}

@article{kjolstad2017tensor,
  title={The tensor algebra compiler},
  author={Kjolstad, Fredrik and Kamil, Shoaib and Chou, Stephen and Lugato, David and Amarasinghe, Saman},
  journal={Proceedings of the ACM on Programming Languages},
  volume={1},
  number={OOPSLA},
  pages={1--29},
  year={2017},
  publisher={ACM New York, NY, USA}
}

@article{senanayake2020sparse,
  title={A sparse iteration space transformation framework for sparse tensor algebra},
  author={Senanayake, Ryan and Hong, Changwan and Wang, Ziheng and Wilson, Amalee and Chou, Stephen and Kamil, Shoaib and Amarasinghe, Saman and Kjolstad, Fredrik},
  journal={Proceedings of the ACM on Programming Languages},
  volume={4},
  number={OOPSLA},
  pages={1--30},
  year={2020},
  publisher={ACM New York, NY, USA}
}

@inproceedings{ahrens2022autoscheduling,
  title={Autoscheduling for sparse tensor algebra with an asymptotic cost model},
  author={Ahrens, Willow and Kjolstad, Fredrik and Amarasinghe, Saman},
  booktitle={Proceedings of the 43rd ACM SIGPLAN International Conference on Programming Language Design and Implementation},
  pages={269--285},
  year={2022}
}

@inproceedings{chen2018tvm,
  title={$\{$TVM$\}$: An automated $\{$End-to-End$\}$ optimizing compiler for deep learning},
  author={Chen, Tianqi and Moreau, Thierry and Jiang, Ziheng and Zheng, Lianmin and Yan, Eddie and Shen, Haichen and Cowan, Meghan and Wang, Leyuan and Hu, Yuwei and Ceze, Luis and others},
  booktitle={13th USENIX Symposium on Operating Systems Design and Implementation (OSDI 18)},
  pages={578--594},
  year={2018}
}

@inproceedings{yadav2022spdistal,
  title={SpDISTAL: Compiling distributed sparse tensor computations},
  author={Yadav, Rohan and Aiken, Alex and Kjolstad, Fredrik},
  booktitle={SC22: International Conference for High Performance Computing, Networking, Storage and Analysis},
  pages={1--15},
  year={2022},
  organization={IEEE}
}

@article{adams2019learning,
  title={Learning to optimize halide with tree search and random programs},
  author={Adams, Andrew and Ma, Karima and Anderson, Luke and Baghdadi, Riyadh and Li, Tzu-Mao and Gharbi, Micha{\"e}l and Steiner, Benoit and Johnson, Steven and Fatahalian, Kayvon and Durand, Fr{\'e}do and others},
  journal={ACM Transactions on Graphics (TOG)},
  volume={38},
  number={4},
  pages={1--12},
  year={2019},
  publisher={ACM New York, NY, USA}
}

@article{chen2018learning,
  title={Learning to optimize tensor programs},
  author={Chen, Tianqi and Zheng, Lianmin and Yan, Eddie and Jiang, Ziheng and Moreau, Thierry and Ceze, Luis and Guestrin, Carlos and Krishnamurthy, Arvind},
  journal={Advances in Neural Information Processing Systems},
  volume={31},
  year={2018}
}

@article{ragan2013halide,
  title={Halide: a language and compiler for optimizing parallelism, locality, and recomputation in image processing pipelines},
  author={Ragan-Kelley, Jonathan and Barnes, Connelly and Adams, Andrew and Paris, Sylvain and Durand, Fr{\'e}do and Amarasinghe, Saman},
  journal={Acm Sigplan Notices},
  volume={48},
  number={6},
  pages={519--530},
  year={2013},
  publisher={ACM New York, NY, USA}
}

@article{mullapudi2016automatically,
  title={Automatically scheduling halide image processing pipelines},
  author={Mullapudi, Ravi Teja and Adams, Andrew and Sharlet, Dillon and Ragan-Kelley, Jonathan and Fatahalian, Kayvon},
  journal={ACM Transactions on Graphics (TOG)},
  volume={35},
  number={4},
  pages={1--11},
  year={2016},
  publisher={ACM New York, NY, USA}
}

@inproceedings{zheng2022amos,
  title={AMOS: enabling automatic mapping for tensor computations on spatial accelerators with hardware abstraction},
  author={Zheng, Size and Chen, Renze and Wei, Anjiang and Jin, Yicheng and Han, Qin and Lu, Liqiang and Wu, Bingyang and Li, Xiuhong and Yan, Shengen and Liang, Yun},
  booktitle={Proceedings of the 49th Annual International Symposium on Computer Architecture},
  pages={874--887},
  year={2022}
}

@inproceedings{zheng2020ansor,
  title={Ansor: Generating $\{$High-Performance$\}$ tensor programs for deep learning},
  author={Zheng, Lianmin and Jia, Chengfan and Sun, Minmin and Wu, Zhao and Yu, Cody Hao and Haj-Ali, Ameer and Wang, Yida and Yang, Jun and Zhuo, Danyang and Sen, Koushik and others},
  booktitle={14th USENIX symposium on operating systems design and implementation (OSDI 20)},
  pages={863--879},
  year={2020}
}

@inproceedings{acharya2018polyhedral,
  title={Polyhedral auto-transformation with no integer linear programming},
  author={Acharya, Aravind and Bondhugula, Uday and Cohen, Albert},
  booktitle={Proceedings of the 39th ACM SIGPLAN Conference on Programming Language Design and Implementation},
  pages={529--542},
  year={2018}
}

@article{tollenaere2023autotuning,
  title={Autotuning convolutions is easier than you think},
  author={Tollenaere, Nicolas and Iooss, Guillaume and Pouget, St{\'e}phane and Brunie, Hugo and Guillon, Christophe and Cohen, Albert and Sadayappan, P and Rastello, Fabrice},
  journal={ACM Transactions on Architecture and Code Optimization},
  volume={20},
  number={2},
  pages={1--24},
  year={2023},
  publisher={ACM New York, NY}
}

@article{charles2005x10,
  title={X10: an object-oriented approach to non-uniform cluster computing},
  author={Charles, Philippe and Grothoff, Christian and Saraswat, Vijay and Donawa, Christopher and Kielstra, Allan and Ebcioglu, Kemal and Von Praun, Christoph and Sarkar, Vivek},
  journal={Acm Sigplan Notices},
  volume={40},
  number={10},
  pages={519--538},
  year={2005},
  publisher={ACM New York, NY, USA}
}

@article{dean2008mapreduce,
  title={MapReduce: simplified data processing on large clusters},
  author={Dean, Jeffrey and Ghemawat, Sanjay},
  journal={Communications of the ACM},
  volume={51},
  number={1},
  pages={107--113},
  year={2008},
  publisher={ACM New York, NY, USA}
}

@article{zaharia2010spark,
  title={Spark: Cluster computing with working sets.},
  author={Zaharia, Matei and Chowdhury, Mosharaf and Franklin, Michael J and Shenker, Scott and Stoica, Ion and others},
  journal={HotCloud},
  volume={10},
  number={10-10},
  pages={95},
  year={2010},
  publisher={Boston, MA}
}

@article{harris2020array,
  title={Array programming with NumPy},
  author={Harris, Charles R and Millman, K Jarrod and Van Der Walt, St{\'e}fan J and Gommers, Ralf and Virtanen, Pauli and Cournapeau, David and Wieser, Eric and Taylor, Julian and Berg, Sebastian and Smith, Nathaniel J and others},
  journal={nature},
  volume={585},
  number={7825},
  pages={357--362},
  year={2020},
  publisher={Nature Publishing Group UK London}
}

@inproceedings{lin1993zpl,
  title={ZPL: An array sublanguage},
  author={Lin, Calvin and Snyder, Lawrence},
  booktitle={International Workshop on Languages and Compilers for Parallel Computing},
  pages={96--114},
  year={1993},
  organization={Springer}
}

@inproceedings{deitz2004abstractionszpl,
  title={Abstractions for dynamic data distribution},
  author={Deitz, Steven J and Chamberlain, Bradford L and Snyder, Lawrence},
  booktitle={Ninth International Workshop on High-Level Parallel Programming Models and Supportive Environments, 2004. Proceedings.},
  pages={42--51},
  year={2004},
  organization={IEEE}
}

%

\setcounter{figure}{0}
\renewcommand{\thefigure}{A\arabic{figure}}
\setcounter{table}{0}
\renewcommand{\thetable}{A\arabic{table}}
\setcounter{algocf}{0}
\renewcommand{\thealgocf}{A\arabic{algocf}}

\newpage
\appendix

\section{Supplementary Material}
\label{sec:appendix}

\subsection{Example of Cannon's Algorithm}
\label{subsec:cannon}

Cannon's algorithm computes $A = B \cdot C$ by arranging all processors in a 2D grid, and assigning a tile of each matrix to each of the processors. Every processor $p$ alternates between systolic shifts of the tiles of $B$ and $C$ with $p$'s row and column neighbors and accumulating the product of $p$'s current $B$ and $C$ tiles into $A$.

We illustrate the task-based pseudocode for Cannon’s algorithm, adapted from code generated by DISTAL~\cite{yadav2022distal}. The program in \Cref{fig:regentcannon} partitions the input matrices into tiles and launches the \CodeIn{tiles} task across the processor grid. Each \CodeIn{tiles} task, in turn, launches multiple \CodeIn{systolic} tasks that fetch tiles of $B$ and $C$ and accumulate their products into the corresponding tile of $A$. This program builds a task graph for distributed matrix multiplication, where data dependencies between tasks are automatically managed by the runtime system. The task space of the \CodeIn{tiles} task is two-dimensional, defined as $(i, j) \in \{0, \ldots, p_x - 1\} \times \{0, \ldots, p_y - 1\}$, where each point corresponds to a tile of the output matrix. The mapping from this task space to the processor space is specified by the mapper declared in the \CodeIn{IndexTaskMap} statement shown in \Cref{fig:cmapper}.


\begin{figure*}[h]
  \centering
  \begin{subfigure}[t]{0.47\textwidth}
  \lstset{language=Regent,basicstyle=\ttfamily\tiny,numbers=left,escapeinside={(*@}{@*)},numbersep=4pt,frame=single,keywordstyle=\color{keywordcolor},commentstyle=\color{commentcolor}}
\begin{lstlisting}
task systolic(A: region(double),
              B: region(double),
              C: region(double))
  writes(A), reads(A, B, C):
  # Internally use cuBLAS for single-GPU matmul.
  A += matmul(B, C)

task tiles(A: region(double), 
           B: region(double), 
           C: region(double),
           pB: partition(B),
           pC: partition(C),
           px: int, i: int, j: int)
  writes(A), reads(A, B, C):
  for k in (0, px):
    k' = (k + i + j) % px
    systolic(A, pB[i, k'], pC[k', j])(*@\label{line:cannon-systolic}@*)

task cannons_matmul(A: region(double), 
                    B: region(double), 
                    C: region(double),
                    px : int, py : int)
  writes(A), reads(A, B, C):
  pA, pB, pC = ... # Partition into px * py tiles.
  for i, j in ({0, 0}, {px, py}):
    tiles(pA[i, j], B, C, pB, pC, px, i, j)
\end{lstlisting}
  \caption{Application: pseudocode for Cannon's algorithm.}
  \label{fig:regentcannon}
  \end{subfigure}
  \hfill
  \begin{subfigure}[t]{0.47\textwidth}
  \lstset{language=Python,basicstyle=\ttfamily\tiny,numbers=left,escapeinside={(*@}{@*)},numbersep=4pt,morekeywords={Task,Region,Layout,Backpressure,GarbageCollect,IndexTaskMap,SingleTaskMap,merge,split,balance_split,decompose,Machine,IPoint,ISpace,MSpace},deletekeywords={is},commentstyle=\color{commentcolor},frame=single,keywordstyle=\color{keywordcolor}}
\begin{lstlisting}
# Define a block mapping for a 2D launch domain.
def block(Task task): (*@\label{line:block}@*)
    # ispace represents 2D launch domain of a x b.
    is = task.ispace (*@\label{line:ispace}@*)
    # ipoint is an index point within launch domain.
    ip = task.ipoint (*@\label{line:ipoint}@*)
    # Get all GPUs as m nodes x n GPUs grid.
    machine = Machine(GPU)
    # Tile nodes; 3-dim grid: (m1 x m2) x n.
    m3d = machine.decompose(0, is) (*@\label{line:autosplitfirst}@*)
    # Tile GPUs; 4-dim grid: (m1 x m2) x (n1 x n2).
    m4d = m3d.decompose(2, is / m3d[:-1]) (*@\label{line:autosplitsecond}@*)
    # Project the index point onto the 4D machine.
    upper = (ip[0] * m4d[0] / is[0], (*@\label{line:uppertuple}@*)
             ip[1] * m4d[1] / is[1])
    lower = (ip[0] % m4d[2], ip[1] % m4d[3]) (*@\label{line:cyclicinvoke}@*)
    # Unpack two 2D tuples to index 4D machine.
    return m4d[*upper, *lower] (*@\label{line:return6d}@*)

IndexTaskMap tiles block (*@\label{line:indextaskmaptask5}@*)

# Map all regions of all tasks to GPU framebuffer.
Region * * GPU FBMEM (*@\label{line:memory}@*)
# Data layout in memory.
Layout * * * C_order (*@\label{line:layout}@*)
# Collect read-only regions used by task_4.
GarbageCollect systolic * (*@\label{line:collectmemory}@*)
# At most 1 systolic task is active at a time.
Backpressure systolic 1 (*@\label{line:instancelimit}@*)

\end{lstlisting}
  \caption{Mapper in \name{} for \cannon algorithm.}
  \label{fig:cmapper}
  \end{subfigure}
\caption{Application code in a task-based programming system and the corresponding \name{} mapper for the \cannon algorithm.}
\label{fig:exampleapp}
\end{figure*}

The mapper that faithfully implements Cannon's algorithm organizes the processors into a 2D grid and blocks the GPUs on each node. Here, we use the \decompose machine transformation twice to block the node and GPU dimensions of the machine to resolve the dimensionality mismatch between the task space and the processor space. Lines \ref{line:memory}--\ref{line:instancelimit} specify the mapping of data to memories. The \CodeIn{Regent} statement specifies that all data should be mapped to GPU framebuffer memory. The \CodeIn{Layout} directive declares that any region used by any task on any processor (denoted by three \CodeIn{*}s) is allocated in row-major (C-style) order, which is necessary to interface with cuBLAS. The \CodeIn{GarbageCollect} directive ensures the read-only tiles of $B$ and $C$ requested by \CodeIn{systolic} tasks are eagerly reclaimed to minimize memory usage. Finally, the \CodeIn{Backpressure} directive limits the number of active \CodeIn{systolic} tasks to 1 per processor to avoid allocating too many tiles at one time.

\subsection{Baseline Greedy Algorithm}
\label{subsec:greedy}

\let\oldnl\nl
\newcommand{\nonl}{\renewcommand{\nl}{\let\nl\oldnl}}

\begin{algorithm}[H]
\small
\caption{Greedy Heuristic for Processor Grid Selection (Suboptimal)}
\label{alg:chapel}
\SetKwFunction{FMain}{Greedy}
\SetKwFunction{FPrime}{PrimeFactorization}
\SetKwFunction{FArgMin}{ArgMin}
\SetKwFunction{FSort}{Sort}
\SetKwProg{Fn}{Function}{:}{}

\nonl\textbf{Input:}\\
\nonl\quad $d$ \ccc{Number of processors}\\
\nonl\quad $k$ \ccc{Dimensionality of the task space}\\

\Fn{\FMain{$d$, $k$}}{
    \texttt{primes} $\leftarrow$ \FPrime($d$) \ccc{Sorted list of prime factors, $d = p_1 \leq \dots \leq p_n$}\label{line:primesort}\\
    \texttt{factors} $\leftarrow$ new int[$k$]\\
    \For{$i \in \{0, \dots, k{-}1\}$}{
        \texttt{factors}[$i$] $\leftarrow 1$
    }
    \For{$p \in$ \texttt{primes}}{
        $j \leftarrow$ \FArgMin(\texttt{factors}) \ccc{Index of smallest current product}\label{line:argmin}\\
        \texttt{factors}[$j$] $\leftarrow$ \texttt{factors}[$j$] $\times\ p$\label{line:multiply}
    }
    \FSort(\texttt{factors}, \texttt{descending=True}) \ccc{Sort for consistent ordering}\label{line:sortagain}\\
    \Return \texttt{factors}
}
\end{algorithm}

\subsection{Generalization of the Decompose Primitive}
\label{subsec:generalization}

While \Cref{sec:decompose} addresses isotropic communication across task space dimensions, the decompose primitive can be extended to anisotropic patterns, such as uneven halo widths and dimension-specific all-to-all exchanges (e.g., for data transposes). This generalization supports a much wider range of computations on block-structured grids. In both cases, only the objective function changes; the same minimization procedure from \Cref{subsec:decompose:impl} still applies.

\subsubsection{Anisotropic halo communication}

To generalize, we extend the communication area from \Cref{subsec:decompose:def} to a volume that captures the total data exchanged between distributed machines. In isotropic cases, where halo width is uniform, volume reduces to area. For anisotropic patterns, varying halo width across dimensions must be considered. Letting $h_n$ denote the width in dimension $n$, the total communication volume $V$ in a $k$-dimensional space is $V = \sum_{n=1}^kd_{i_n}h_n\prod_{m\neq n }l_m$. For $l_n = w_nd_{i_n}$, the volume, $V$, can also be expressed as $V = \left(\sum_{n=1}^k\frac{h_n}{w_n}\right)\left(\prod_{m=1}^k l_m\right)$, where $\prod_{m=1}^k l_m$ is constant for a certain task space.

\subsubsection{Transpose via all-to-all communication along certain dimensions}
For transpose operations via all-to-all communication, the data volume per partition along the $n$-th dimension is given by $v_n = \frac{d_{i_n} - 1}{d_{i_n}} \prod_{m=1}^k w_m$. This reflects that each partition splits its data evenly into $d_{i_n}$ groups, one of which remains local while the others are sent to distinct peers along the $n$-th dimension. Thus, the total communication volume within a pencil of $d_{i_n}$ aligned processors is $d_{i_n} v_n$. To optimize the mapping, we minimize the total communication volume, $\left( V + \sum_{n \in \mathbb{T}} V_n^* \right)$, where $\mathbb{T}$ denotes the set of dimensions requiring transposes. The volume along dimension $n$, $V_n^*$, is defined as:
\[
V_n^* = v_n \prod_{m=1}^k d_{i_m} = \left(1 - \frac{1}{d_{i_n}} \right) \left( \prod_{m=1}^k w_m \right) d_i.
\]

\subsection{Example Mapping Functions in \name{}}
\label{subsec:moreexample}

\begin{figure}[H]
    \centering\includegraphics[keepaspectratio=true,width=0.95\columnwidth]{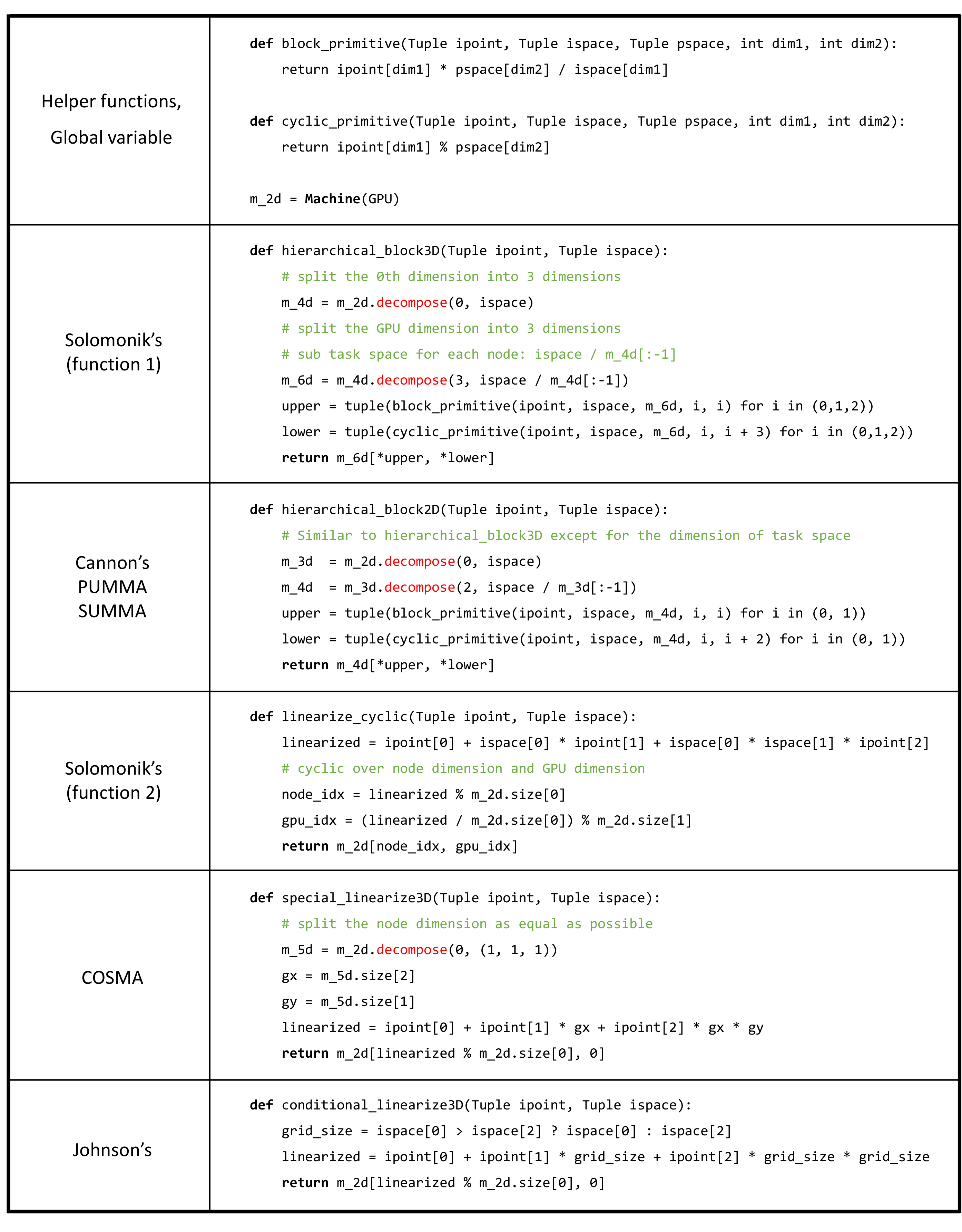}
    \caption{Example functions used by the mappers for the distributed matrix-multiplication algorithms.}
    \label{fig:morefunctions}
\end{figure}

\subsection{Case Studies on Mapper Performance}
\label{subsec:app:perf}

Here we present a case study highlighting the critical role of precise control over mapping strategies. Our investigation demonstrates significant performance variations resulting from minor alterations in the \CodeIn{hierarchical\_block2D} function for the \cannon, \pumma, and \summa algorithms. Specifically, we contrast the mapping function specified by the algorithm with an alternative approach integrating runtime heuristics: the runtime system dynamically assigns the index point to one of the four GPUs on the node, selecting the GPU with the least workload at runtime, rather than adhering to a predetermined distribution.

\begin{figure*}[!tb]
    \centering
    \begin{subfigure}{0.3\textwidth}
        \includegraphics[width=\linewidth]{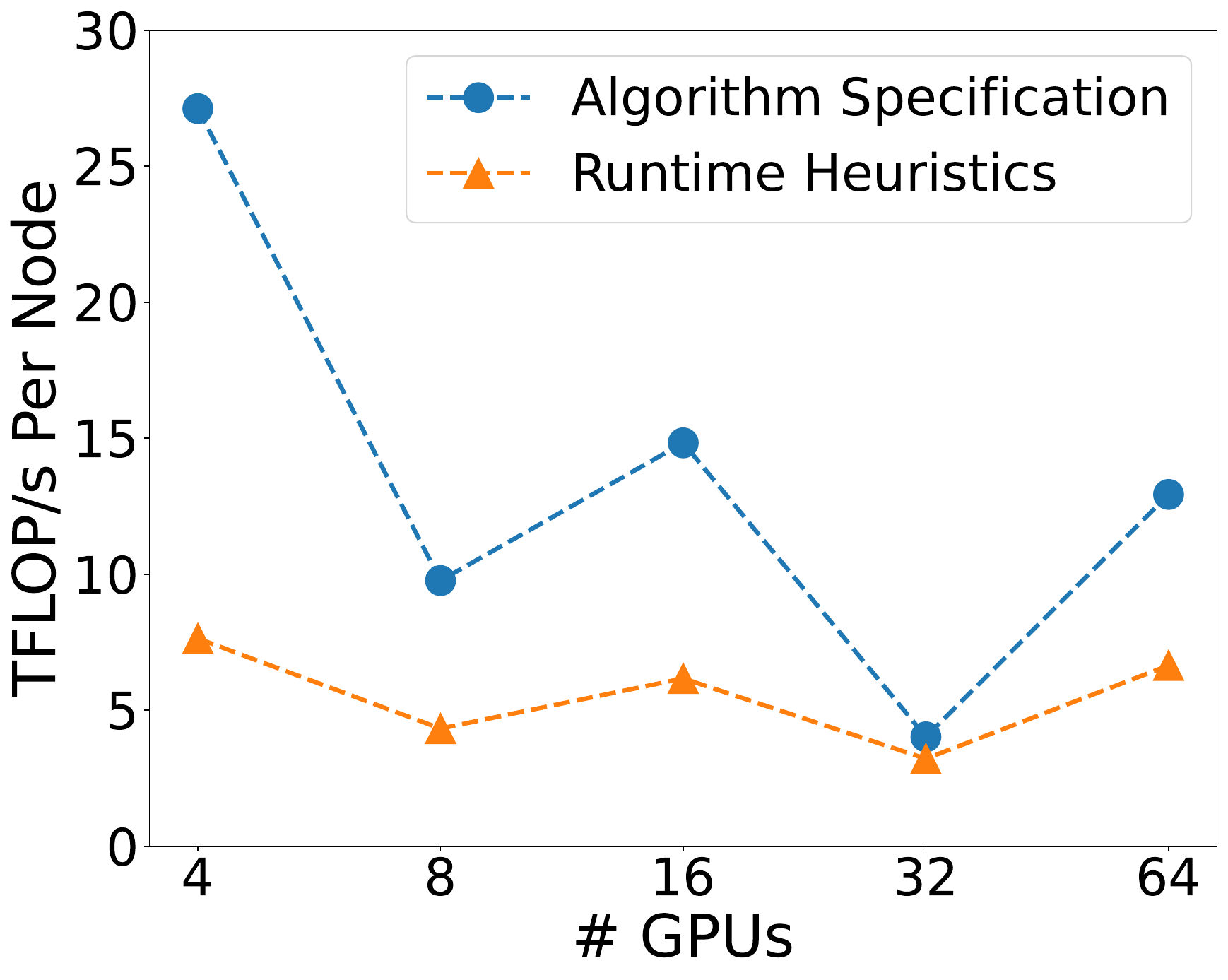}
        \caption{\cannon}\label{fig:cannonperf}
    \end{subfigure}
    \hfill
    \begin{subfigure}{0.3\textwidth}
        \includegraphics[width=\linewidth]{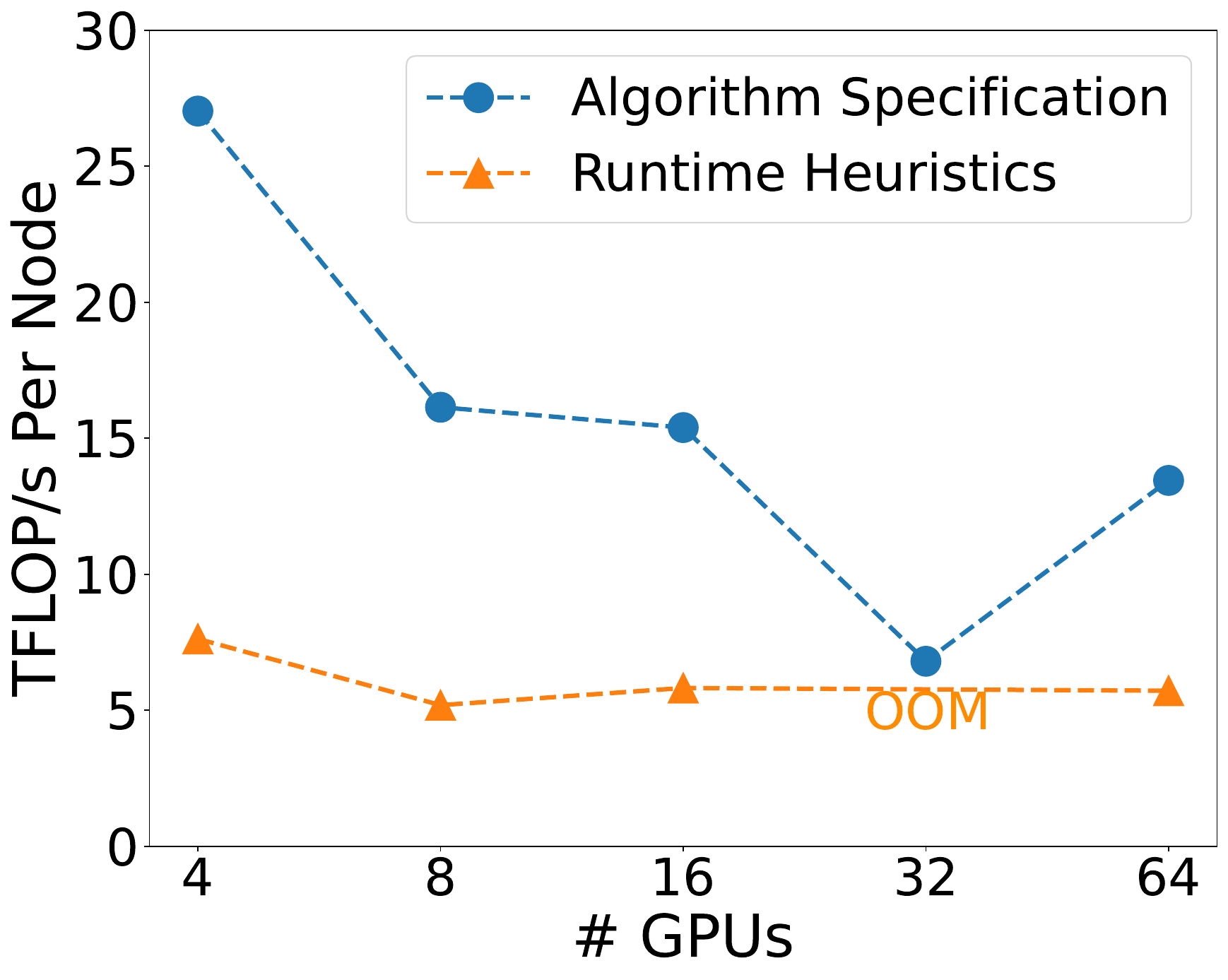}
        \caption{\pumma}\label{fig:pumma}
    \end{subfigure}
    \hfill
    \begin{subfigure}{0.3\textwidth}
        \includegraphics[width=\linewidth]{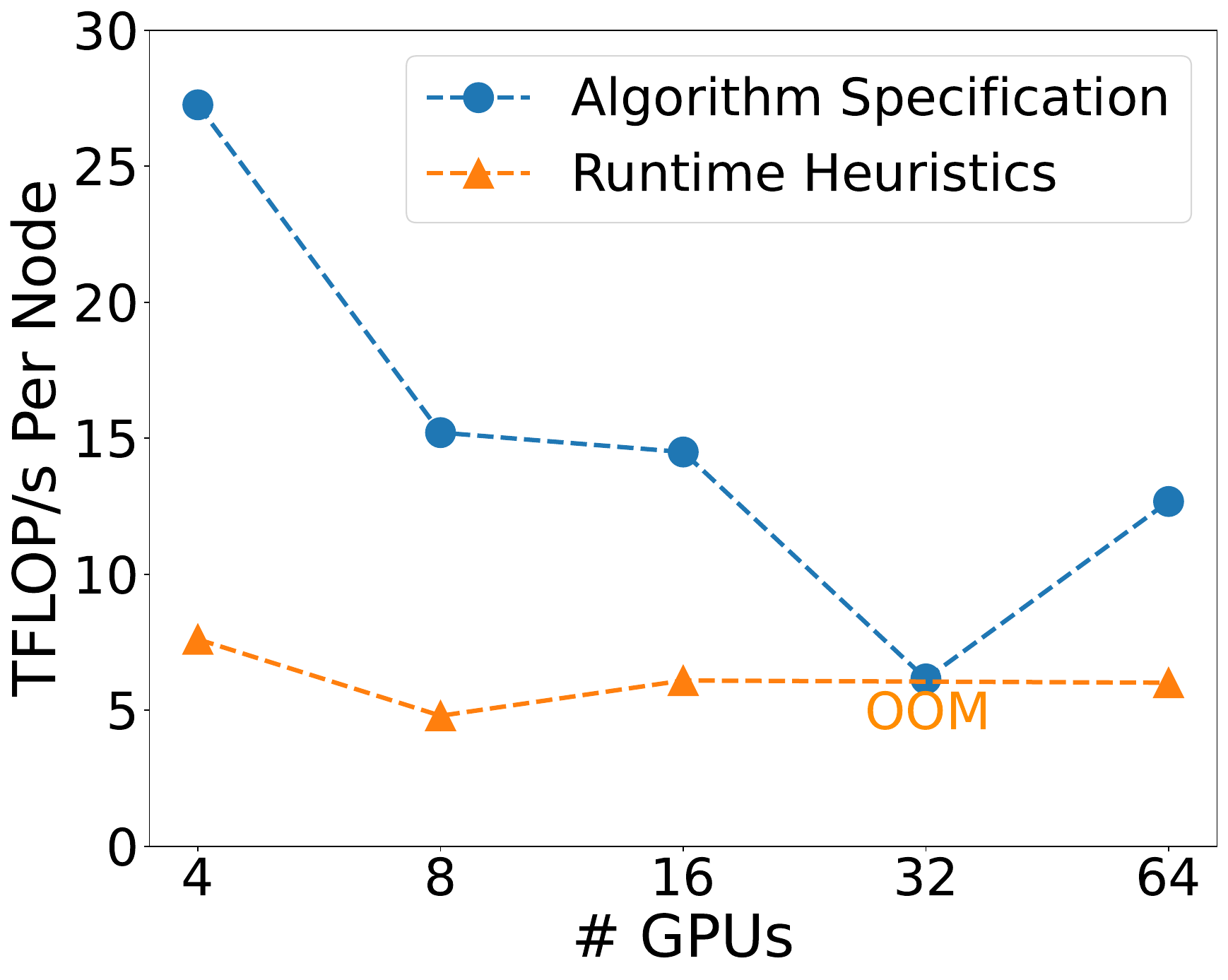}
        \caption{\summa}\label{fig:summa}
    \end{subfigure}
    \precaptionspace
    \caption{Throughput comparison between the original \CodeIn{hierarchical\_block2D} mapping function (specified by the \cannon, \pumma, and \summa algorithm) and a runtime heuristics-based mapper. The runtime heuristics-based mapping can lead to out-of-memory error (denoted by ``OOM'').}
    \label{fig:weakscale}
\end{figure*}

The throughput result depicted in \Cref{fig:weakscale} underscores the pronounced performance discrepancies between the two approaches. The x-axis indicates different machine sizes, and the y-axis indicates the throughput per node. The ``Algorithm Specification'' line shows the throughput achieved by correctly implementing the corresponding algorithms, which match the performance results reported in prior work. The ``Runtime Heuristics'' line shows the throughput achieved by the alternative approach. In the 1-node scenario, the slowdown can reach up to \slowdownmax. This disparity stems from the additional data movement induced by divergent mapping decisions. Furthermore, the runtime heuristics-based mapping leads to out-of-memory (OOM) errors on 32-GPU runs for the \pumma and \summa algorithms, because mapping decisions influence where the data is physically materialized in memory. This example shows that it is important to precisely control the mapping to minimize data movement and optimize memory resource utilization.

\end{document}